\documentclass[aps,pra,notitlepage,reprint,onecolumn,showpacs,preprintnumbers,amsmath,amssymb,superscriptaddress,longbibliography,12pt]{revtex4-1}
\usepackage{graphicx}
\usepackage{bm}
\usepackage{braket}
\usepackage{natbib}
\usepackage{amsmath}

\usepackage{graphicx}
\graphicspath{{pict/}{subm/}{}}
\usepackage{times}
\renewcommand

\usepackage{hyperref}
\hypersetup{pdfstartview={FitH},pdfpagemode={UseNone},
	colorlinks,linkcolor=blue, citecolor=blue, urlcolor=blue,
	bookmarksopen=true, pdfnewwindow=true}
\usepackage[all]{hypcap}

\begin{document}
	\title{Multidimensional synthetic chiral-tube lattices via~nonlinear~frequency~conversion}
	
	\author{Kai Wang}
	\thanks{Present address: Ginzton Laboratory, Stanford University, Stanford, CA 94305, USA}
	\email{wkai@stanford.com}
	\affiliation{Nonlinear Physics Centre, Research School of Physics, The Australian National University, Canberra, ACT 2601, Australia}
	
	\author{Bryn Bell}
    \affiliation{Institute of Photonics and Optical Science (IPOS), School of Physics, University of Sydney, Sydney, NSW 2006, Australia}
	\affiliation{QOLS, Department of Physics, Imperial College London, London SW7 2AZ, UK}
	
	\author{Alexander S. Solntsev}
	\affiliation{School of Mathematical and Physical Sciences, University of Technology Sydney, Ultimo, NSW 2007 Australia}
		\affiliation{Nonlinear Physics Centre, Research School of Physics, The Australian National University, Canberra, ACT 2601, Australia}
	
	\author{Dragomir~N.~Neshev}
	\affiliation{Nonlinear Physics Centre, Research School of Physics, The Australian National University, Canberra, ACT 2601, Australia}
	
	\author{Benjamin J. Eggleton}%
	\affiliation{Institute of Photonics and Optical Science (IPOS), School of Physics, University of Sydney, Sydney, NSW 2006, Australia}
	
	\author{Andrey A. Sukhorukov}
	\email{andrey.sukhorukov@anu.edu.au}
	\affiliation{Nonlinear Physics Centre, Research School of Physics, The Australian National University, Canberra, ACT 2601, Australia}

	\date{\today}
	
	\begin{abstract} 
Geometrical dimensionality plays a fundamentally important role in the topological effects arising in discrete lattices. While direct experiments are limited by three spatial dimensions, the research topic of synthetic dimensions implemented by the frequency degree of freedom in photonics is rapidly advancing. The manipulation of light in such artificial lattices is typically realized through electro-optic modulation, yet their operating bandwidth imposes practical constraints on
the range of interactions between different frequency components.
Here we propose and experimentally realize all-optical synthetic dimensions involving specially tailored simultaneous short- and long-range interactions between discrete spectral lines mediated by frequency conversion in a nonlinear waveguide. We realize triangular chiral-tube lattices in three-dimensional space and explore their  four-dimensional generalization.
We implement a synthetic gauge field with nonzero magnetic flux and observe the associated multidimensional dynamics of frequency combs, all within one physical spatial port. We anticipate that our method will provide a new means for the fundamental study of high-dimensional physics and
act as an important step towards using 
topological effects in optical devices operating in the time and frequency domains.
	\end{abstract}
	
	\maketitle
	
\newpage

\section*{Introduction}
	
Discrete photonic lattices constitute a versatile platform for topological photonics~\cite{Lu:2014-821:NPHOT}, with various implementations employing arrays of evanescently coupled waveguides~\cite{Christodoulides:2003-817:NAT}, metamaterials~\cite{Engheta:2006:Metamaterials}, and coupled resonators~\cite{Mittal:2014-87403:PRL}. In these systems, the most commonly considered topological features originate from the dispersion associated with the wave-vector space, where accordingly the geometrical dimensionality fundamentally limits the degrees of freedom that can contribute to the topological invariant. As such, a possibility to access higher geometrical dimensions is a key enabling factor for drastically boosting the manifestations of topological effects. This motivates the rapidly developing field of synthetic dimensions~\cite{Yuan:2018-1396:OPT,Ozawa:2019-349:NRP}, where many schemes to artificially create extra dimensions were proposed~\cite{Boada:2012-133001:PRL,Celi:2014-43001:PRL,Yuan:2016-741:OL,Ozawa:2016-43827:PRA,Martin:2017-41008:PRX} and experimentally demonstrated~\cite{Regensburger:2012-167:NAT, Stuhl:2015-1514:SCI, Mancini:2015-1510:SCI, Lustig:2019-356:NAT, Maczewsky:2020-76:NPHOT}. Generally, higher dimensionality is equivalent to increased connectivity, thus a multidimensional lattice can be synthesized by lower- or even one-dimensional (1D) lattices with long-range couplings extending beyond the nearest-neighbors~\cite{Casanova:2012-190502:PRL, Grass:2015-63612:PRA, Yuan:2018-104105:PRB}. 
Importantly, the higher-dimensional formalism can reveal extra nontrivial geometrical and topological properties incorporated 
into the original 1D wave vector space. Therefore, an essential yet challenging task to facilitate topological photonics in synthetic dimensions~\cite{Lustig:2019-356:NAT} lies in the development of artificial lattices exhibiting exotic topological behavior enabled by effectively larger dimensionality, such as higher-order topological modes~\cite{Dutt:1911.11310:ARXIV}. 

In topological photonics, beyond the consideration of dimensionality there are other essential ingredients contributing to the versatile topological phenomena. Of key importance is the boundary condition that has been extensively studied in the well-known bulk-boundary correspondence~\cite{Yuan:2016-741:OL, Ozawa:2016-43827:PRA, Lustig:2019-356:NAT}. While these are usually considered for lattices with edges, the periodic boundary conditions in  extended lattice are also physically relevant. For instance, in carbon nanotubes, the specific way that the honeycomb lattice gets wrapped into the tube can dramatically impact on the material properties, yet the associated topological characteristics remain largely unexplored. Beyond the most familiar types of zigzag and armchair carbon nanotubes, the most general situation arises in between these two cases corresponding to a \emph{chiral} periodic boundary condition, where chiral-tube lattices are formed~\cite{Artyukhov:2014-4892:NCOM}. Notably, synthetic photonic systems provide diverse and flexible platforms to artificially arrange such lattices, going beyond the natural material arrangements. In particular, it is of fundamental interest to explore various lattice types in addition to honeycomb and analyze a possibility to realize multidimensional analogues of chiral tube structures.

Another important ingredient of topological photonics lies in the gauge potential and the associated gauge field, due to their fundamental role in describing the movement of charged particles. The realization of artificial gauge fields in photonics has underpinned novel manifestations of light emulating charged-particle dynamics, facilitating the multifaceted aspects of topological photonics such as breaking of time-reversal symmetry~\cite{Mittal:2014-87403:PRL,Bell:2017-1433:OPT,Qin:2018-133901:PRL,Lustig:2019-356:NAT,Yuan:2016-741:OL}, non-reciprocal light guiding~\cite{Lumer:2019-339:NPHOT,Lin:2014-31031:PRX}, and topological lasing~\cite{Harari:2018-eaar4003:SCI, Bandres:2018-eaar4005:SCI}. In a tight-binding photonic lattice, in general, an artificial gauge field corresponds to different phases acquired by light when it couples from site A to site B compared to the opposite path from B to A. 
Such gauge fields can give rise to a flux associated with encircling a geometrical area in 2D or higher-dimensional generalizations, which is essential to many topological models, such as the Landau gauge~\cite{Yuan:2016-741:OL, Dutt:2020-59:SCI} with globally non-zero flux and the Haldane model~\cite{Haldane:1988-2015:PRL,Yuan:2018-104105:PRB} with locally non-zero flux.

Temporal and spectral behaviors of light play an important role in many research fields and applications, from telecommunications to the spectroscopy of materials. Importantly, many shortcomings restricting the performance of photonic systems arise in the temporal domain, such as group-velocity
dispersion. 
It is thereby important to achieve topological effects for robust operation in temporal or spectral systems in a regime compatible with common applications, such as frequency combs generated from CMOS-compatible integrated resonators with a free spectral range (FSR) in the GHz regime~\cite{Razzari:2010-41:NPHOT}. 
So far, the control of spectral couplings was primarily studied~\cite{Yuan:2016-741:OL,Ozawa:2016-43827:PRA,Yuan:2018-104105:PRB} and realized~\cite{Dutt:2019-162:ACSP, Dutt:2019-3122:NCOM, Reimer:1909.01303:ARXIV, Dutt:2020-59:SCI} based on electro-optic modulation (EOM). However, this approach fundamentally 
limits the frequency separation between coupled modes by the EOM bandwidth, which commonly restricts the induced coupling to the nearest spectral lines in synthetic frequency lattices. 
In contrast, all-optical approaches based on parametric nonlinearity~\cite{Bersch:2009-2372:OL,Bell:2017-1433:OPT} and photon-phonon interaction~\cite{Kang:2009-276:NPHYS,Wolff:2017-23021:NJP,Eggleton:2019-664:NPHOT} appear as promising solutions to bring multidimensional topological photonics to devices requiring ultra-fast temporal modulation and accordingly large FSR. 
Recently, we reported the implementation of synthetic long-range coupling in an all-optical system mediated by parametric nonlinearity within one spatial mode (port) with up to 100~GHz separation between the spectral lines~\cite{Bell:2017-1433:OPT}.
However, a possibility to realize multidimensional synthetic lattices in an all-optical platform 
so far remained unexplored.

In this work, for the first time to our knowledge, we theoretically establish and experimentally demonstrate that all-optical spectral lattices 
can synthesize multidimensional chiral lattices 
in combination with nontrivial gauge fields.
The synthetic dimensions are based on simultaneous specially tailored short- and long-range couplings between discrete frequency components, which are mediated by optical nonlinearity and directly controlled by the spectral shape of the optical pump.
With three orders of coupling, we show the construction of triangular chiral-tube lattices in three dimensions and establish their four-dimensional generalization. We also develop a pump configuration that induces nontrivial artificial gauge fields associated with effective nonzero magnetic flux, 
and experimentally demonstrate their influence on
a quantum walk in triangular chiral-tube lattices. 

We note that our all-optical implementation achieves broad operating bandwidth of hundreds of GHz, which exceeds the capabilities of complex electro-optic modulation schemes considered previously while offering greater simplicity. The high bandwidth enables a direct matching with the free-spectral range of integrated resonators, which can facilitate multiple application of synthetic multidimensional lattices presented in our work.
Moreover, a capability for multidimensional and coherent reshaping of discrete frequency lines can enable unconventional and non-reciprocal manipulation of quantum frequency combs~\cite{Reimer:2016-1176:SCI}, which may boost the capacity of photonic quantum communications and information processing.

	\section*{Results}

\subsection*{Construction of synthetic dimensions in a nonlinear waveguide}
	
\begin{figure}[tb]
\centering
\includegraphics[width=1\columnwidth]{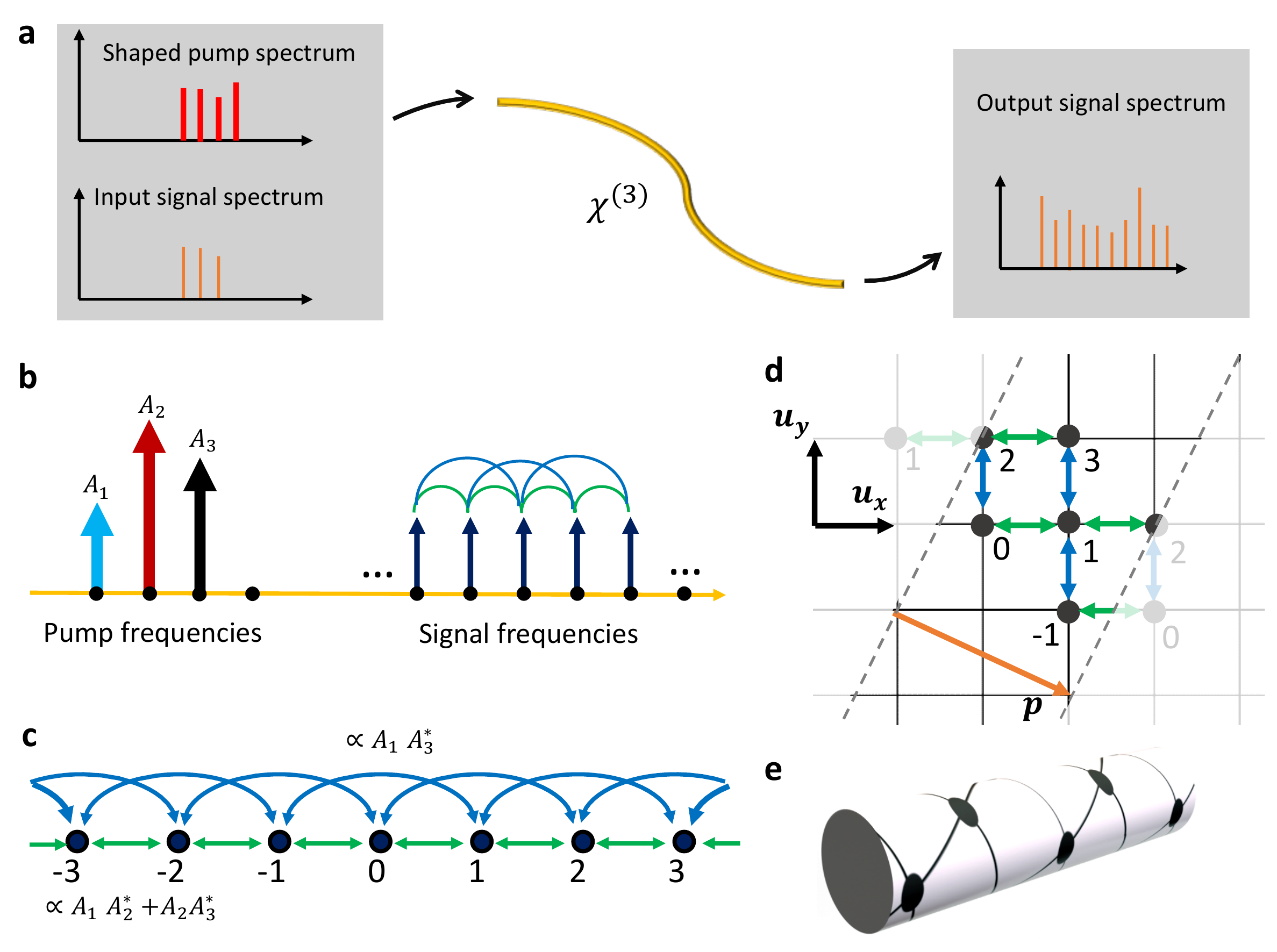}
\caption{{\bf Conceptual sketch of constructing multidimensional synthetic lattices in a nonlinear fiber.} {\bf a,}~A nonlinear waveguide with $\chi^{(3)}$ nonlinearity, where a shaped pump mediates the conservative interactions between signal frequencies, giving rise to a reshaped signal spectrum at the output. {\bf b,}~An example of pump profile that induces cross-talk between one and two unit frequency separations. {\bf c,}~The corresponding spectral lattice with first- ($C_1$) and second-order ($C_2$) couplings driven by the nonlinear interactions mediated by a shaped pump spectrum shown in ({\bf b}). {\bf d,}~Synthetic two-dimensional square lattice constructed using the spectral lattice in ({\bf c}). {\bf e,} Illustration of the chiral tube lattice formed by wrapping the lattice in ({\bf d}) with the chiral periodic boundary condition $\mathbf{p}$. }\label{fig1}
\end{figure}

We start by introducing a general approach for the implementation of spectral photonic lattices in a nonlinear waveguide. As sketched in Fig.~\ref{fig1}{\bf a}, a spectral lattice is realized by using a shaped pump composed of several equidistant frequencies with the spacing $\Omega$. Such a pump can drive the interactions between the frequency components on the input signal spectrum all inside one fiber or waveguide with $\chi^{(3)}$ nonlinearity, in the regime of the so-called four-wave-mixing Bragg scattering (FWMBS)~\cite{Bell:2017-1433:OPT}. Under energy conservation and undepleted pump approximation,
each pair of pump frequencies separated by $n\Omega$ drives the 
coupling between two signal lines with the same frequency difference $n\Omega$, as shown in Fig.~\ref{fig1}{\bf b}. Such discrete spectral lines form a lattice (Fig.~\ref{fig1}{\bf c}), where each frequency represents one lattice site. Importantly, nonlocal and complex-valued couplings can be implemented by specially tailoring the pump spectrum~\cite{Bell:2017-1433:OPT}. 

Now we outline the key concept of exact mapping between higher-dimensional lattices and a one-dimension spectral lattice with nonlocal couplings induced through nonlinear frequency conversion.
In the example shown above in Figs.~\ref{fig1}{\bf a--c}, three pumps equally separated by $\Omega$
introduce coupling of first and second orders~(Fig.~\ref{fig1}{\bf c}).
Then, the evolution of the signal spectrum along the nonlinear waveguide in the phase-matching regime is governed by the Hamiltonian in terms of the creation ($\hat{a}_m^\dagger$) and annihilation ($\hat{a}_m$) operators for the discrete signal frequency components, 
\begin{equation}\label{eq:FWM}
	\mathbf{H}=-\sum_{m}  \sum_{\{n\}} C_{n} \hat{a}_{m}^\dagger \hat{a}_{m+n}-\mathrm{H.c.},
\end{equation}
where a set of positive integers $\{n\}$ indicates the orders of coupling and $C_n$ are the corresponding coupling constants of the $n$-th order. The `H.c.' denotes Hermitian conjugate, and $m$ is an integer running through all phase-matched spectral lines. The coupling constants $C_n$ are given by the following expression~\cite{Bell:2017-1433:OPT},
\begin{equation} \label{eq:Cn_pump}
    C_n = 2\gamma P \sum_m A_m A^\ast_{m-n},
\end{equation}
where $\gamma$ is the effective nonlinearity, and $P$ is the average pump power. Here $A_m$ denotes the complex amplitudes of pump spectral components in the fiber, which are normalized as $\sum_m |A_m|^2=1$. The evolution of the wavefunction governed by the Hamiltonian in Eq.~\eqref{eq:FWM} can be expressed as %
\begin{equation} \label{eq:psi_H}
   \ket{\psi(z)}=\exp (i z \mathbf{H}) = \exp (i z P \mathbf{H'}), 
\end{equation}
where $z$ is the propagation distance along the fiber, and we denote by $\mathbf{H'}=\mathbf{H} / P$ a normalized Hamiltonian that is independent on the total pump power. We see that the wavefunction dynamics can be 
observed by varying the average pump power $P$ for a fixed fiber length $z=L$, such that 
$P$ effectively acts as time.
	
We now consider a nontrivial and representative case of two coupling orders $n=1,2$, and show how the spectral lattice is mapped to a two-dimensional square lattice. The general idea is based on the mapping of each specific order of coupling to a certain basis vector in higher-dimensional space. For the example shown in Fig.~\ref{fig1}{\bf d}, which is a two-dimensional space of square lattice, there are two basis vectors, $\mathbf{u_x}$ and $\mathbf{u_y}$. Hence we can map the coupling order $n=1$ to $\mathbf{u_x}$, and $n=2$ to $\mathbf{u_y}$. Then, we obtain a Hamiltonian in the two-dimensional space that represents a square synthetic lattice
\begin{equation}\label{eq:sq}
\mathbf{H_{sq}}=-\sum_{m} \left[  C_{1} \hat{a}_{\mathbf{r_m}}^\dagger \hat{a}_{\mathbf{r_m}+\mathbf{u_x}}+C_{2} \hat{a}_{\mathbf{r_m}}^\dagger \hat{a}_{\mathbf{r_m}+\mathbf{u_y}} \right] -\mathrm{H.c.},
\end{equation}
where $\mathbf{r_m}$ is a vector indicating the spatial coordinate of the $m$-th site in this two-dimensional space. 
To provide an exact mapping, it is essential to reflect in 2D the algebraic property of the one-dimensional lattice, where a sequence of two first-order couplings produces the same  frequency shift $2 \Omega$ as a second-order coupling. This property can be satisfied by imposing a periodic boundary condition for the two-dimensional synthetic space, as shown in Fig.~\ref{fig1}{\bf d}, where the orange arrow represents the wrapping vector $\mathbf{p}=2\mathbf{u_x}-\mathbf{u_y}$.
Consequently, the two-dimensional equivalent lattice is actually wrapped into a $(2,-1)$ chiral tube connected by the dashed lines in Fig.~\ref{fig1}{\bf d}. Such a chiral tube is schematically illustrated in Fig.~\ref{fig1}{\bf e}.

\subsection*{Observation of a quantum walk in synthetic triangular lattices}

\begin{figure}[t]
		\centering
		\includegraphics[width=1\columnwidth]{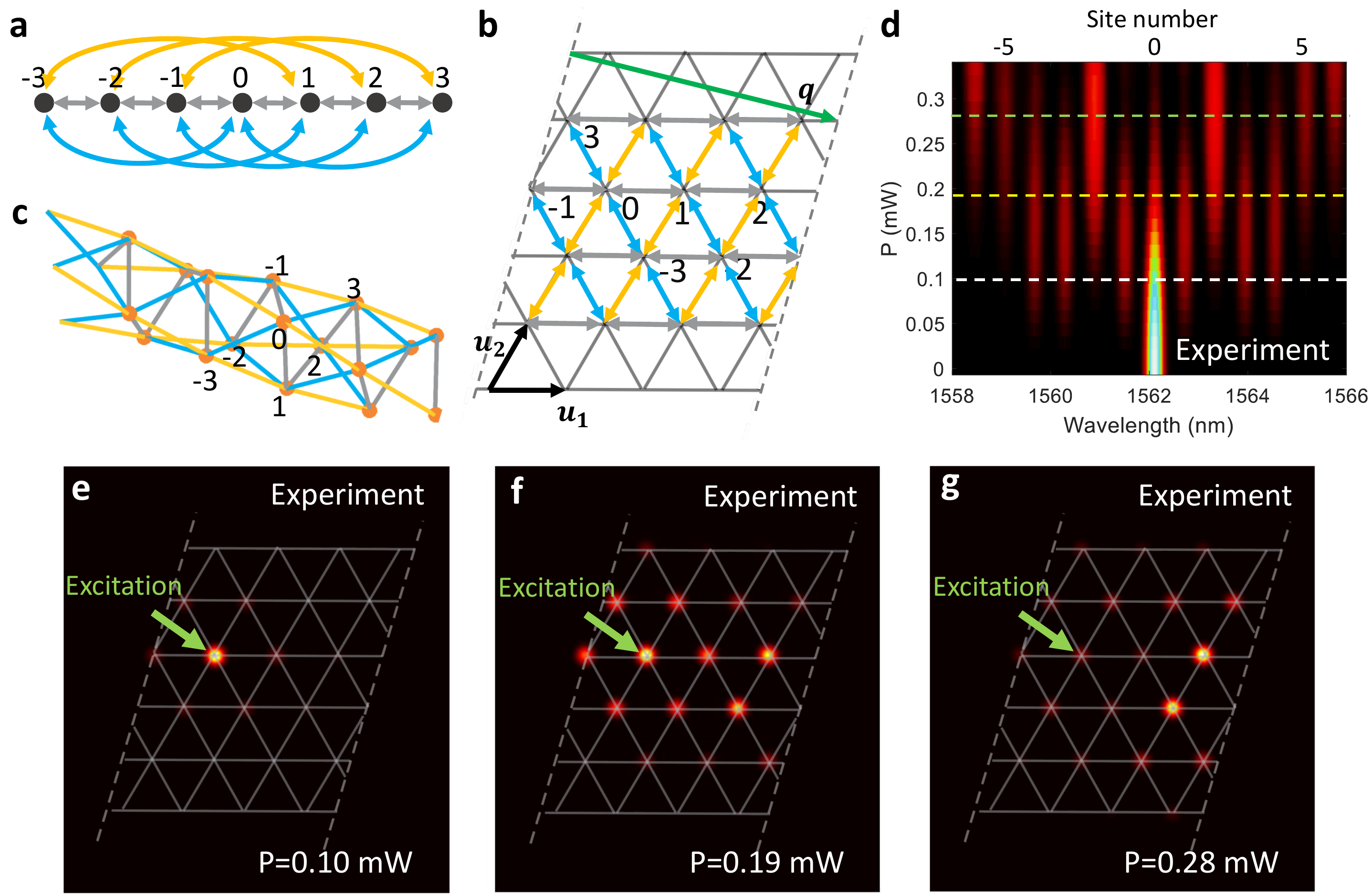}
		\caption{{\bf Experimental observation of quantum walk in a synthetic two-dimensional triangular chiral-tube lattice.} {\bf a,} A lattice with three coupling orders 1, 3 and 4, where only couplings to the shown sites are plotted with arrows. {\bf b,}~The corresponding synthetic triangular lattice in two dimensional space. {\bf c,}~3D sketch of the lattice in ({\bf b}). {\bf d,}~Experimental realization of a frequency quantum walk, where $P$ is the power at each of the three pump spectral components with $A_1=A_2=A_5$. {\bf e--g,} Mapping of experimental data from ({\bf d}) to a two-dimensional triangular lattice with $P=0.10,\ 0.19,\ 0.28$~mW as indicated by labels. }\label{fig2}
	\end{figure} 

We formulate and demonstrate experimentally an original mapping procedure for the realization of a synthetic triangular lattice. This presents a nontrivial case with non-orthogonal basis vectors, which
has not been considered on any platform in previous studies. We show that a triangular lattice can be obtained by mapping from a spectral lattice with specially engineered 
simultaneously short- and long-range couplings.
The synthetic frequency space is sketched in Fig.~\ref{fig2}{\bf a}, where 1st, 3rd and 4th orders of coupling are present.
In the two-dimensional space of triangular lattice, as shown in Fig.~\ref{fig2}{\bf b}, the basis vectors are $\mathbf{u_1}=\left[1,\ 0 \right]^{\mathrm{T}}$ and $\mathbf{u_2}=\left[1/2,\ \sqrt{3}/2 \right]^{\mathrm{T}}$, which are not orthogonal with each other. We map the 1st order coupling to the vector $\mathbf{u_1}$, and the 4th order coupling to $\mathbf{u_2}$. Then, we find that the 3rd order coupling is automatically mapped to $\mathbf{u_3}=\mathbf{u_2}-\mathbf{u_1}$. This constructs the two-dimensional equivalent triangular lattice as sketched in Fig.~\ref{fig2}{\bf b}, with the Hamiltonian
\begin{equation}
\mathbf{H_{tr}}=-\sum_{m} \left[  C_{1} \hat{a}_{\mathbf{r_m}}^\dagger \hat{a}_{\mathbf{r_m}+\mathbf{u_1}}+C_{3} \hat{a}_{\mathbf{r_m}}^\dagger \hat{a}_{\mathbf{r_m}+\mathbf{u_3}}+C_{4} \hat{a}_{\mathbf{r_m}}^\dagger \hat{a}_{\mathbf{r_m}+\mathbf{u_2}} \right] - \mathrm{H.c.}.
\end{equation}
Similar to the example of the square lattice discussed above (Fig.~\ref{fig1}{\bf d}), there appears a periodic boundary condition. It is defined by the wrapping vector
$\mathbf{q}=4\mathbf{u_1}-\mathbf{u_2},$
shown as a green arrow in Fig.~\ref{fig2}{\bf b}. Hence the triangular lattice is effectively wrapped and connected by the dashed lines in Fig.~\ref{fig2}{\bf b}. To show this more intuitively, we sketch a 3D visualisation of the $(4,-1)$ chirally wrapped tube in Fig.~\ref{fig2}{\bf c}. We determine the unit cell vector of the tube lattice as $\mathbf{l}=-2\mathbf{u_1}+7\mathbf{u_2}$ (not shown in the figure), which is the shortest vector that can connect two sites along the parallel direction of the tube ($\mathbf{l}^{\mathrm{T}}\mathbf{q}=0$ due to orthogonality).

We now present an experimental realization of quantum walk in the multidimensional synthetic lattice space. Quantum walks have been observed in various types of photonic lattices using classical laser sources, where the evolution of coherent light is mathematically analogous to the quantum single-particle dynamics~\cite{Perets:2008-170506:PRL}. We tailor the complex amplitudes $A_m$ of the pump spectral lines to induce the desired couplings in the signal frequency lattice according to Eq.~\eqref{eq:Cn_pump}.
Specifically, we employ three pumps with equal amplitudes $A_1=A_2=A_5$ to achieve the frequency lattice as illustrated in Fig.~\ref{fig2}{\bf a} with equal 1st, 3rd and 4th order couplings, $C_1=C_3=C_4$. We shape the pump with no phase difference between the complex amplitudes at different frequencies, and therefore all couplings are real-valued. With a single-frequency signal excitation, we observe a quantum walk in this frequency space, as shown in Fig.~\ref{fig2}{\bf d}.  
We map this experimentally-realized synthetic lattice to the triangular lattice as outlined in Figs.~\ref{fig2}{\bf b,c}. The mapped quantum walk is shown in Figs.~\ref{fig2}{\bf e--g} at three representative average pump powers $P=0.10,\ 0.19, \ 0.28$ mW, respectively. As mentioned above, here pump power acts as time in the quantum walk. In these figures, the site of excitation is marked by a green arrow. This represents 
an experimental observation of quantum walk in higher synthetic dimensions. Our results agree quite well with the corresponding theoretical predictions calculated by coupled mode equations. An animated image illustrating the dynamics of the experiment (incorporating  Figs.~\ref{fig2}{\bf e--g}) and comparison to theory is provided as Figure S1 in Supplementary files. 

\begin{figure}[t]
		\centering
		\includegraphics[width=0.9\columnwidth]{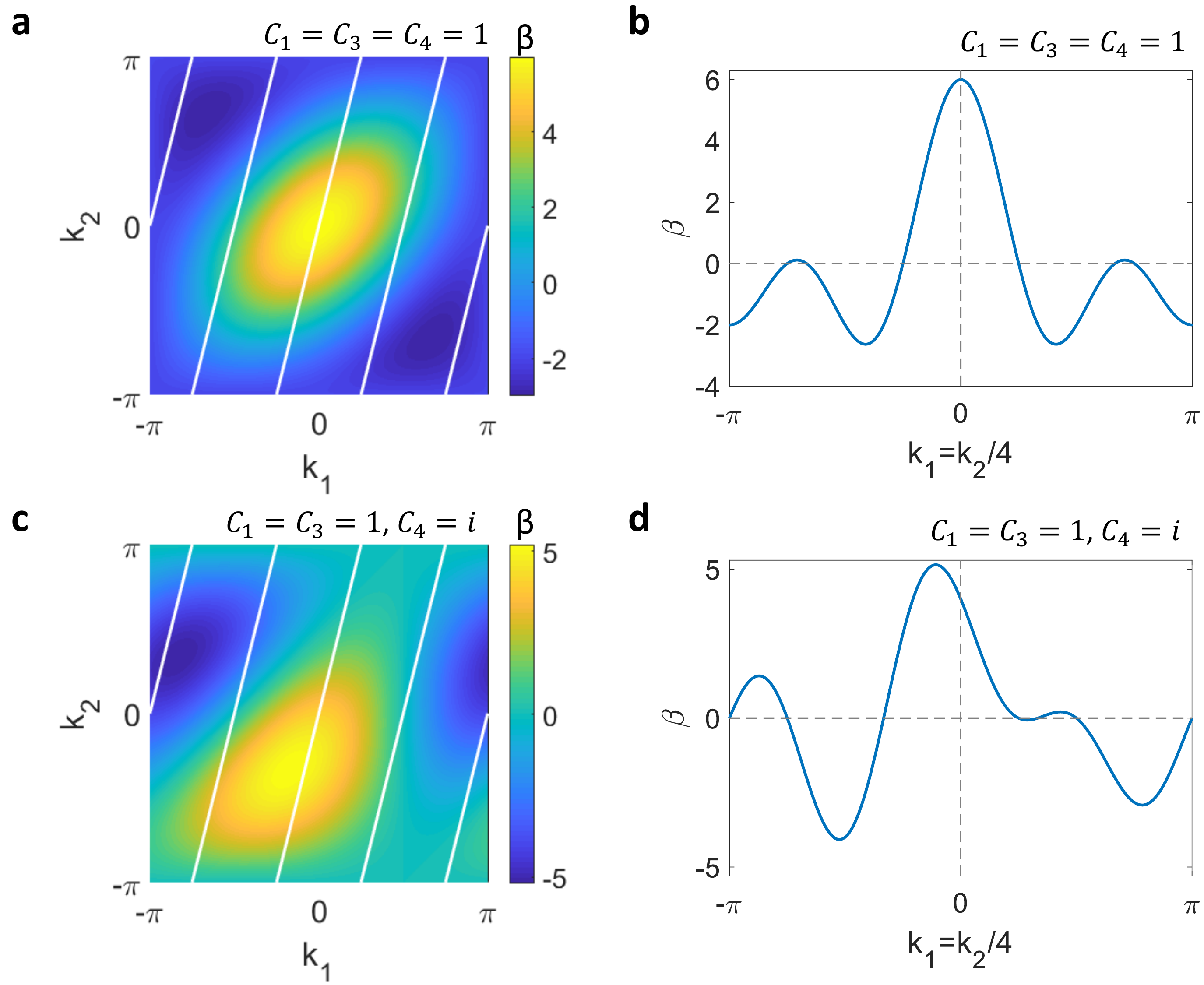}
		\caption{{\bf Dispersion of the triangular lattice wrapped to a $(4,-1)$ chiral tube.} {\bf a,} Color density plot of the propagation constant $\beta$ vs. wave numbers $k_1$ and $k_2$, where the $(4,-1)$ chiral periodic boundary condition only allows $k_1,k_2$ values on the white lines. Here we set $C_1=C_3=C_4=1$.  {\bf b,}~The corresponding dispersion of ({\bf a}) plotted in 1D along $k_1=k_2/4$. {\bf c,d,}~Analogous plots as ({\bf a,b}) with a complex coupling $C_4=i$, other parameters are the same. }\label{fig:disp}
	\end{figure}

To provide an insight into the properties of our mapped synthetic chiral tube lattice, we also perform a theoretical analysis of the wave dispersion. 
We apply the Bloch theorem and calculate the propagation constant as $\beta(k_1,k_2)=2 \mathrm{Re}[C_1 \exp{(ik_1)}+C_4 \exp{(ik_2)}+C_3 \exp{(ik_3)}]$, where $k_1$, $k_2$, $k_3$ are the wave numbers in the reciprocal space of basis vectors $\mathbf{u_1},\mathbf{u_2},\mathbf{u_3}$, respectively, and $k_3 \equiv k_2-k_1$.  
In Fig.~\ref{fig:disp}{\bf a} we plot a representative case for $C_1=C_3=C_4=1$. Due to the periodic boundary condition, not all values of $k_1,k_2$ are allowed. For the $(4,-1)$ chiral tube discussed above we have $4k_1-k_2=2N\pi$, where $N$ is an integer. Such allowed values are denoted as white lines in Fig.~\ref{fig:disp}{\bf a}. We trace out $k_1=k_2/4$ for the range of $-\pi$ to $\pi$ and present the dispersion as a 1D curve plotted in Fig.~\ref{fig:disp}{\bf b}. For comparison, we also show another case with a different coupling $C_4=i$ in Fig.~\ref{fig:disp}{\bf c}. We see that the resulting 1D dispersion shown in Fig.~\ref{fig:disp}{\bf d} becomes asymmetric due to the complex coupling $C_4$ that breaks the time-reversal symmetry. We show in the following section that this regime is associated with the appearance of gauge field in the mapped high-dimensional lattices.

\subsection*{Artificial gauge field with non-zero magnetic flux in chiral-tube lattices}

\begin{figure}[b]
	\centering
	\includegraphics[width=1\columnwidth]{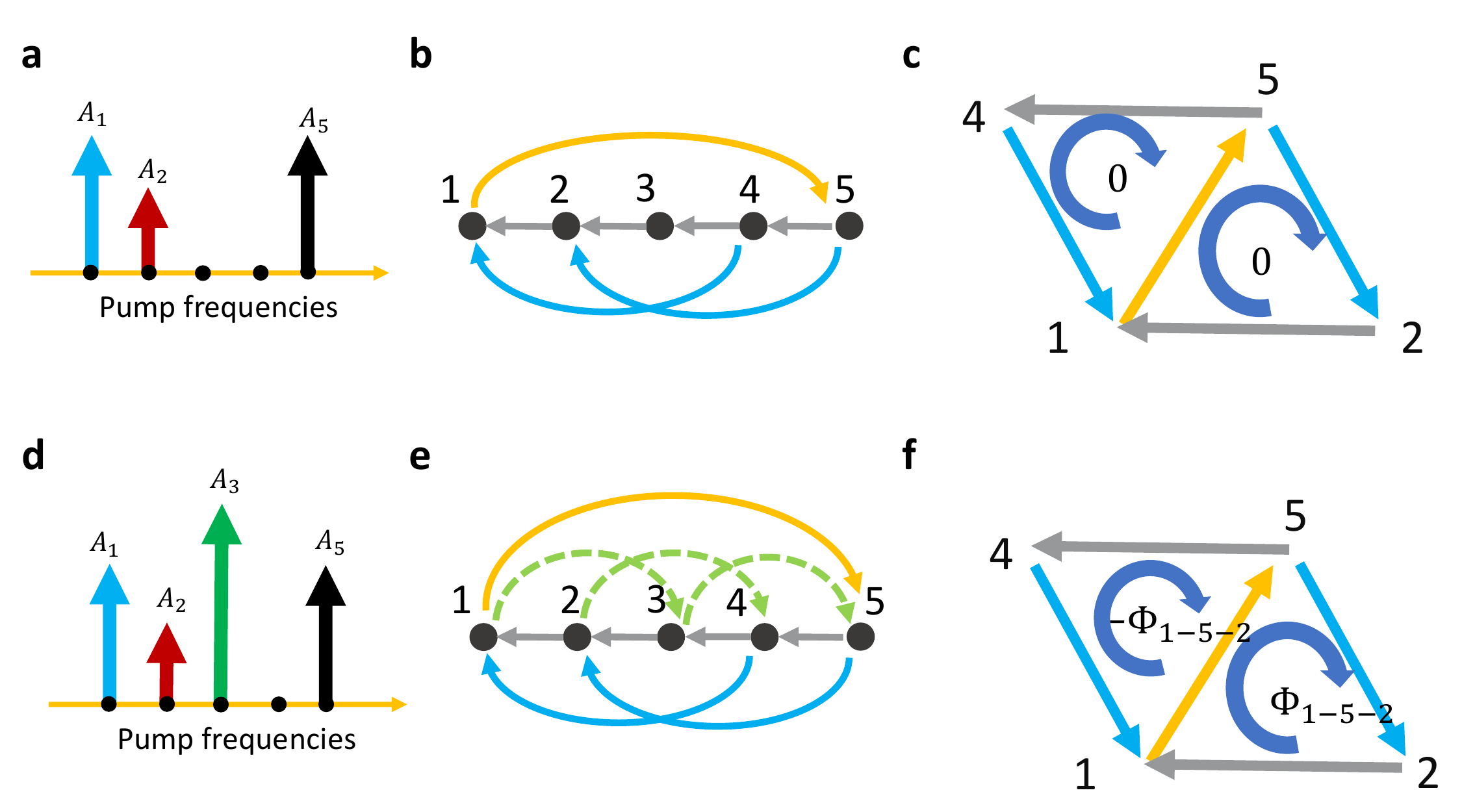}
	\caption{ {\bf Implementation of nontrivial gauge field in synthetic triangular lattices.} {\bf a,} Three-pump configuration to induce three orders of coupling. {\bf b,} The implemented synthetic frequency lattice with 1st, 3rd, and 4th orders of coupling. {\bf c,} The mapping of ({\bf b}) to a two-dimensional triangular lattice where the flux in each cell is zero. {\bf d--f,} Our specially designed scheme for implementing nonzero flux by adding an extra pump ($A_3$) and tailoring the pump phases. }\label{fig:sp_md_gauge}
\end{figure} 

We demonstrate that the complex-valued nature of the coupling constants in the synthetic frequency lattice can enable the artificially created gauge fields. In contrast to the one-dimensional case, in higher-dimensional lattices an important aspect of the gauge potential is associated with the induced magnetic flux, which can arise in presence of non-zero phase accumulation around a closed loop in the lattice. 

To illustrate the capacity of our scheme to synthesize a nontrivial gauge field, we first revisit the mapping of a spectral lattice with three orders of coupling as sketched in Fig.~\ref{fig2}{\bf a}. To induce such couplings, the minimum number of pumps is three with complex amplitudes $A_1$, $A_2$, and $A_5$, as illustrated in Fig.~\ref{fig:sp_md_gauge}{\bf a}, with the corresponding couplings $C_1 \propto A_2 A_1^\ast$, $C_3 \propto A_5 A_2^\ast$, and $C_4 \propto A_5 A_1^\ast$.
In Fig.~\ref{fig:sp_md_gauge}{\bf b} we show a section of a spectral lattice implemented by the pump configuration in Fig.~\ref{fig:sp_md_gauge}{\bf a}, visualizing five sites (1 to 5) as an illustration. 
We use a one-way arrow to show each order of coupling, where the coupling to the other direction simply takes the complex-conjugate value due to Hermiticity. We determine the phases of each order of the coupling along the direction of the arrows in Fig.~\ref{fig:sp_md_gauge}{\bf b} as
$\phi_{1}=\arg (A_2)-\arg (A_1)$ (gray arrow), $\phi_{3}=\arg (A_5)-\arg (A_2)$ (blue arrow), and $\phi_{-4}=\arg (A_1)-\arg (A_5)$ (orange arrow). This lattice is mapped to the triangular lattice using the approach described above, which is shown in Fig.~\ref{fig:sp_md_gauge}{\bf c} for the first five sites. We find that the clockwise flux vanishes in each of the triangular cells,
\begin{equation}\label{eq:zeroflux}
\begin{split}
\Phi_{1-5-2}&=\phi_{-4}+\phi_{3}+\phi_{1}=0,\\
\Phi_{1-4-5}&=-\phi_{-4}-\phi_{3}-\phi_{1}=0.
\end{split}
\end{equation}
This shows that with the minimum necessary number of three pumps, the number of free parameters is not sufficient to implement nonzero flux in any of the triangular cells of this two-dimensional lattice. 

We reveal that a nonzero flux can be induced by adding an extra pump, indicated with the green arrow in Fig.~\ref{fig:sp_md_gauge}{\bf d} with amplitude $A_3$. The corresponding couplings between the signal frequencies are sketched in Fig.~\ref{fig:sp_md_gauge}{\bf e}. Since we are still aiming for coupling orders 1,3,4, first we need to ensure that the second-order coupling, denoted by the green dashed arrows in Fig.~\ref{fig:sp_md_gauge}{\bf e}, is cancelled out,
\begin{equation}\label{eq:c2}
C_2 \propto A_3 A_1^\ast+A_5 A_3^\ast=0.
\end{equation}
As a sufficient condition to fulfill  Eq.~\eqref{eq:c2}, in our experiment we take $|A_1|=|A_5|$, $\arg(A_3 A_1^\ast)=\pi/4$, and $\arg(A_5 A_3^\ast)=-3\pi/4$. Then, we calculate the phases of the other orders of coupling, and find that $\phi_{3}$ (blue arrow) and $\phi_{-4}$ (orange arrow) remain the same as in the case analyzed above with three pumps. This happens since each of the two orders is induced by the same pair of pumps.
Importantly, the first order coupling acquires a different phase according to the expression
\begin{equation}
C_1\propto A_2 A_1^\ast+A_3 A_2^\ast.
\end{equation}
This allows us to implement an arbitrary phase $\phi_1 = \arg(C_1)$ in experiment, where we fix $\arg(A_3 A_1^\ast)=\pi/4$ but freely choose the amplitudes of all the three involved pumps and the phase of $A_2$.
Therefore, the limitation given in Eqs.~\eqref{eq:zeroflux} no longer applies and we can engineer any nonzero flux $\Phi_{1-5-2}$. Note that the following condition still holds
\begin{equation}
\Phi_{1-4-5}\equiv-\Phi_{1-5-2},
\end{equation}
which leads to a zero flux if one encircles a pair of neighboring cells. This is an analogous situation to the Haldane model~\cite{Haldane:1988-2015:PRL}, where the total flux over all cells is zero yet locally there appear locations with nonzero flux.

\begin{figure}[tb]
	\centering
	\includegraphics[width=\columnwidth]{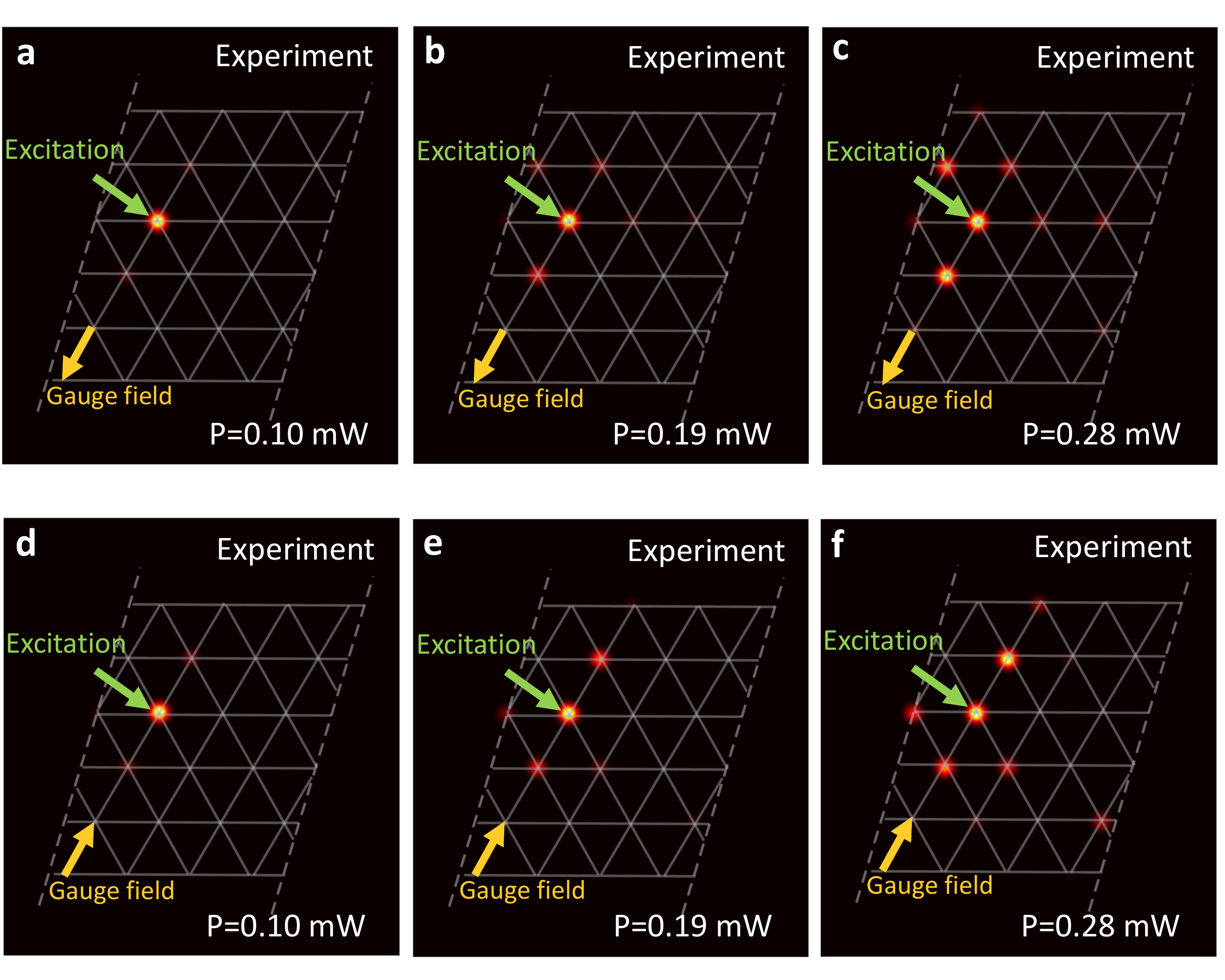}
	\caption{{\bf Experimental observation of quantum walks in synthetic dimension with  artificial gauge fields.} {\bf a--c,} Experimental results for an effective complex coupling phase $\pi/2$ along the direction of the yellow arrow, for pump average powers $P=0.10,\ 0.19,\ 0.28$ mW, respectively. {\bf d--f,} The corresponding cases of ({\bf a--c}) with an opposite direction of the $\pi/2$ phase. }\label{fig:sp_md_walk2}
\end{figure} 

Next, we show a representative set of experimental results that demonstrate such nontrivial gauge potentials. We intentionally make $|C_4|$ slightly larger than $|C_1|=|C_3|$ in order to more clearly observe the features associated with the artificial gauge potential. Specifically, we choose the pump profiles with four frequencies to obtain the coupling constants of $C_1:C_3:C_4=3:-3:5 \exp (i \alpha)$. This arrangement effectively corresponds to a phase of $\pi-\alpha$ along the $\mathbf{u_2}$ basis vector (positive direction), if we have a gauge transformation to make all couplings other than those along $\mathbf{u_2}$ real-valued. In two sets of experiments presented in Fig.~\ref{fig:sp_md_walk2}, we realize quantum walks with a single-site excitation in the synthetic triangular lattice with $\alpha=-\pi/2$ and $\alpha=\pi/2$.
In Figs.~\ref{fig:sp_md_walk2}{\bf a--c} we show the case with $\alpha=-\pi/2$, where the yellow arrow indicates the direction along which there is a positive $\pi/2$ phase in the coupling.
For the gradually increasing pump powers as indicated in Figs.~\ref{fig:sp_md_walk2}{\bf a--c}, we find that the evolution of the single-site excitation exhibits an asymmetric behavior along the direction of the effective gauge field. This can be clearly seen by comparing to the case $\alpha=\pi/2$, which is shown in Figs.~\ref{fig:sp_md_walk2}{\bf d--f} for the same pump powers as in {\bf a--c}, respectively. In particular, the patterns formed in the quantum walk as shown in Figs.~\ref{fig:sp_md_walk2}{\bf c} and {\bf f} look like two arrows pointing in the opposite directions. Comparisons of the experimental results with theory for Figs.~\ref{fig:sp_md_walk2}{\bf a--c} and {\bf d--f} are shown using two animated images in the supplementary files as Figure~S2 and Figure~S3, respectively. 
	
\subsection*{Higher-dimensional analogues of tube lattices}
\begin{figure}[t]
		\centering
		\includegraphics[width=1\columnwidth]{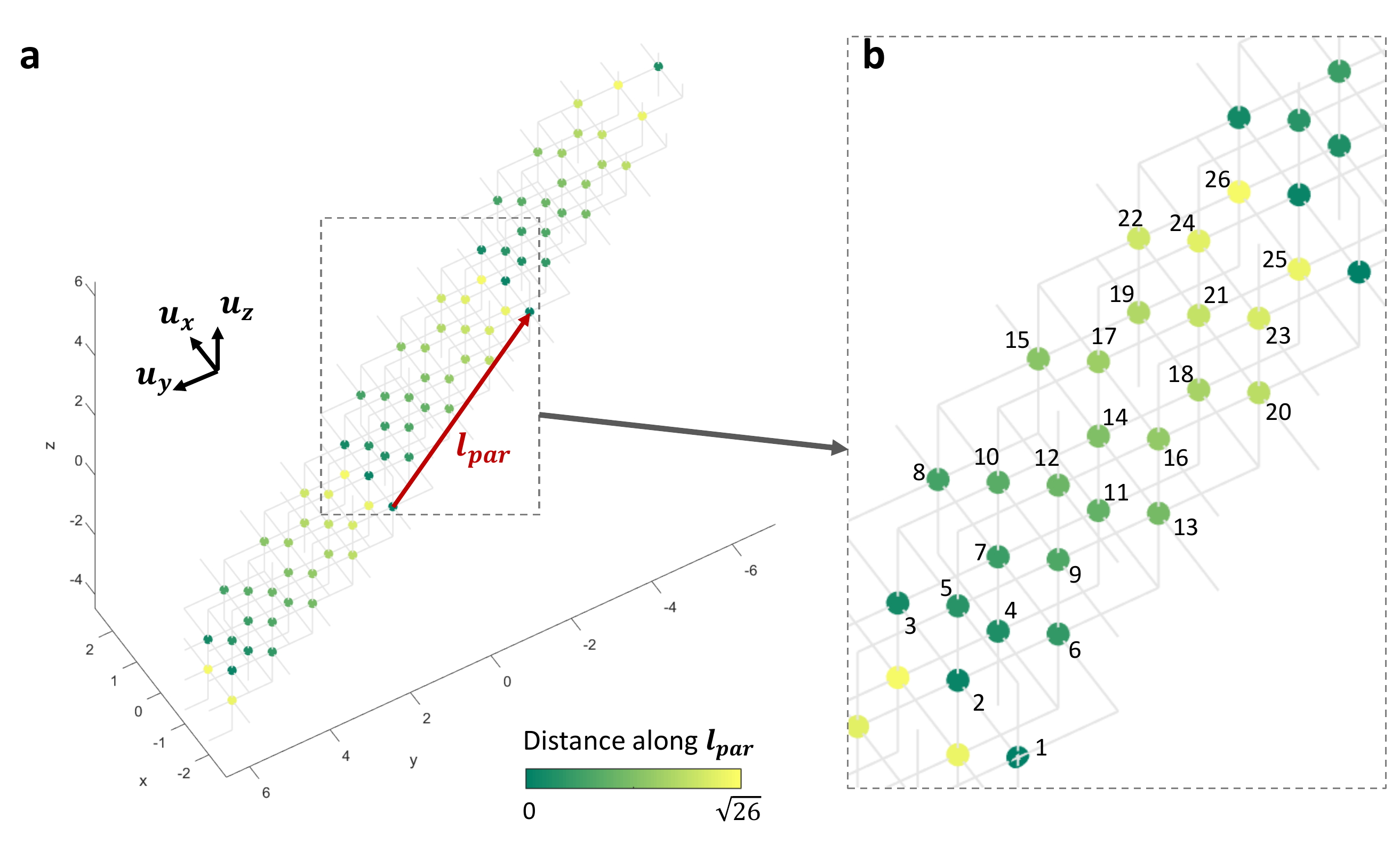}
		\caption{{\bf Construction of a four-dimensional analogue of chiral tube lattices.} {\bf a,}~Sites of a cubic lattice constructed from the coupling orders 1, 3 and 4, which is wrapped in a fourth dimension forming a chiral lattice.  {\bf b,}~Zoomed-in view of the dashed panel in ({\bf a}), with the sites in one unit cell numbered from 1 to 26. }\label{fig:4d}
	\end{figure}

We now discuss how three-dimensional (3D) cubic lattices can be constructed through mapping and how the periodic boundary conditions wrap the cubic lattice into a four-dimensional (4D) analogue of the 3D chiral tubes considered above. We keep using the lattice configuration in Fig.~\ref{fig2}{\bf a} as an example, which involves coupling orders 1, 3 and 4,  however we perform a different mapping procedure. We map the 1st order coupling to the basis vector $\mathbf{u_x}$, the 3rd order coupling to $\mathbf{u_z}$, and the 4th order to $-\mathbf{u_y}$, see Fig.~\ref{fig:4d}{\bf a}. By doing so, we effectively realize the Hamiltonian
\begin{equation}\label{eq:cubic}
\mathbf{H_{cub}}=-\sum_{m} \left[  C_{1} \hat{a}_{\mathbf{r_m}}^\dagger \hat{a}_{\mathbf{r_m}+\mathbf{u_x}}+C_{4} \hat{a}_{\mathbf{r_m}}^\dagger \hat{a}_{\mathbf{r_m}-\mathbf{u_y}}+C_{3} \hat{a}_{\mathbf{r_m}}^\dagger \hat{a}_{\mathbf{r_m}+\mathbf{u_z}} \right] -\mathrm{H.c.}.
\end{equation}

The mapped cubic lattice is subject to a nontrivial periodic boundary condition.
In contrast to the 2D case where the periodic boundary is described by a wrapping vector, as discussed above, here the condition can be expressed as a wrapping plane $\mathbf{s}$, given by the equation $x-4y+3z=\mathrm{const}$. Within each wrapping plane, we keep lattice sites with no repetitions, which gives rise to the lattice structure shown in Fig.~\ref{fig:4d}{\bf a}. We determine the unit cell vector as $\mathbf{l_{par}}=\mathbf{u_x}-4\mathbf{u_y}+3\mathbf{u_z}$, which is shown as a red arrow in Fig.~\ref{fig:4d}{\bf a}. The color map in Fig.~\ref{fig:4d}{\bf a} shows the coordinate of each lattice site in a unit cell along $\mathbf{l_{par}}$. We further zoom in the dashed panel in Fig.~\ref{fig:4d}{\bf a} and show it as Fig.~\ref{fig:4d}{\bf b}, where we number all the 26 sites in one unit cell. Note that although this lattice structure is visualized in 3D, there are couplings (connections) enabled by a fourth dimension analogous to the wrapping of a tube. For example, in this 3D layout, site 15 only has two neighboring sites, i.e. 14 and 19 (see Fig.~\ref{fig:4d}{\bf b}); however, it actually also interacts with sites 11, 12, 16, 18, 19 (connections not shown) enabled by a fourth dimension. This is the first example of using synthetic lattice to realize a 4D generalization of tube lattices. We note that the geometry of such a structure is equivalent to an open 3-torus~\cite{Thurston:1997:ThreeDimensional}, which is one of the important models used to study the topology of the universe~\cite{Weeks:2001:ShapeSpace}.
	
	\section*{Discussion}

To summarize, we have theoretically constructed and experimentally realized nonlinearity-induced synthetic frequency lattices, in which mapping forms multidimensional chiral-tube lattices. The working principle based on nonlinear frequency conversion enables an all-optical realization with large separation between the spectral lines, overcoming the bandwidth limitations of systems employing electro-optic modulation. We observed quantum walks in artificial two-dimensional triangular lattices wrapped into a 3D tube and implemented gauge fields with nonzero flux. We also showed the construction of a 4D analogue of tube lattices employing the chiral periodical boundary conditions formed in a mapped 3D cubic lattice. We point out that the periodic boundary condition, which is an important ingredient of this work, can also be suppressed if so desired by using a combination of large and incommensurate coupling orders. 
It is also an interesting open question on how to develop a general mathematical formalism that establishes a mapping between 1D lattices with arbitrary coupling range and higher-dimensional lattices with different lattice types.
We anticipate that our general conceptual approach may be implemented in a variety of optical setups including electro-optic modulation (EOM) and can also stimulate new realizations of lattices in the spatial domain~\cite{Eichelkraut:2014-268:OPT, Lustig:2019-356:NAT, Maczewsky:2020-76:NPHOT}. Furthermore, synthetic lattices can enable new applications for single-shot reconstruction of the amplitude, phase and coherence of signal spectra \cite{Wang:2002.09160:ARXIV}.

We further note that by employing the process of sum-frequency-generation mediated by the second-order nonlinearity, one could implement two sub-lattices that realize a synthetic honeycomb lattice. The onsite potential of each lattice site would be determined by the phase mismatch, which can enable the implementation of edges by dispersion engineering of the nonlinear waveguide, opening a path to  
the exploration of topological properties associated with edge-boundary correspondence. Parametric nonlinearity can also be used to realize gain and induce 
synthetic lattices 
with non-Hermitian topological properties.
Importantly, our approach is in general non-reciprocal as phase-matching condition is fulfilled in one direction of the nonlinear waveguide determined by the pump~\cite{Wang:2017-1990:OL}, yet such a system is free of the limitations imposed by dynamic reciprocity~\cite{Shi:2015-388:NPHOT} associated with nonlinear devices. Additionally, since our approach is mediated by optical nonlinearity, it is naturally suitable for the exploration of nonlinear effects in synthetic space, such as multidimensional solitons~\cite{Jukic:2013-13814:PRA}. 
Future work may also consider developing schemes to implement other types of synthetic gauge fields in synthetic space, such as those corresponding to a uniform magnetic field that can induce a circular motion of wavepackets in photonics \cite{Fang:2012-782:NPHOT, Longhi:2015-2941:OL}.
Above all, we anticipate that our work can boost new fundamental advances and practical experiments in multidimensional topological, nonlinear and quantum photonics.

\section*{MATERIALS AND METHODS}
\subsection*{Experimental setup for a synthetic frequency lattice in a nonlinear fiber}

The nonlinear frequency conversion was realized in a highly nonlinear fiber with a length of 750 m, the zero-dispersion wavelength of which is 1551 nm. The coherent light source is a mode-locked laser with approximately 25 nm bandwidth, which was reshaped by two spectral wave shapers (Finisar WaveShaper 4000S), following the approach of Ref.~\cite{Bell:2017-1433:OPT}. In the first wave shaper, the laser was split into two channels, with one as a pump channel and the other as a signal channel. The pump channel was then launched into an Erbium-doped fiber amplifier (EDFA) for amplification, followed by a variable attenuator to change the average pump power. The signal channel went through a tunable delay line, which could be adjusted for the signals to match the pump in time. The two channels were then recombined using the second waveshaper, which also shaped the required signal profile (as input state of the spectral lattice) and removed the noise induced by the EDFA to the pump. The output spectra were observed with an optical spectrum analyzer.

\section*{CONFLICT OF INTEREST}
The authors declare no conflict of interest.

\section*{AUTHOR CONTRIBUTIONS}
K.W., A.S.S., D.N.N., and A.A.S. developed the theoretical concept and performed numerical modelling. B.B. and B.J.E. developed the experimental setup and performed the measurements.
B.J.E. and A.A.S. supervised the project. All authors analysed the results and co-wrote the paper.
	
\section*{Acknowledgments}
		
		We gratefully acknowledge financial support from Australian Research Council: Discovery Project (DP160100619, DP190100277); Centre of Excellence CUDOS (CE110001018); Laureate Fellowship (FL120100029). We also thank Daniel Leykam for useful dicussions.   


\begin{thebibliography}{48}%
\makeatletter
\providecommand \@ifxundefined [1]{%
 \@ifx{#1\undefined}
}%
\providecommand \@ifnum [1]{%
 \ifnum #1\expandafter \@firstoftwo
 \else \expandafter \@secondoftwo
 \fi
}%
\providecommand \@ifx [1]{%
 \ifx #1\expandafter \@firstoftwo
 \else \expandafter \@secondoftwo
 \fi
}%
\providecommand \natexlab [1]{#1}%
\providecommand \enquote  [1]{``#1''}%
\providecommand \bibnamefont  [1]{#1}%
\providecommand \bibfnamefont [1]{#1}%
\providecommand \citenamefont [1]{#1}%
\providecommand \href@noop [0]{\@secondoftwo}%
\providecommand \href [0]{\begingroup \@sanitize@url \@href}%
\providecommand \@href[1]{\@@startlink{#1}\@@href}%
\providecommand \@@href[1]{\endgroup#1\@@endlink}%
\providecommand \@sanitize@url [0]{\catcode `\\12\catcode `\$12\catcode
  `\&12\catcode `\#12\catcode `\^12\catcode `\_12\catcode `\%12\relax}%
\providecommand \@@startlink[1]{}%
\providecommand \@@endlink[0]{}%
\providecommand \url  [0]{\begingroup\@sanitize@url \@url }%
\providecommand \@url [1]{\endgroup\@href {#1}{\urlprefix }}%
\providecommand \urlprefix  [0]{URL }%
\providecommand \Eprint [0]{\href }%
\providecommand \doibase [0]{https://doi.org/}%
\providecommand \selectlanguage [0]{\@gobble}%
\providecommand \bibinfo  [0]{\@secondoftwo}%
\providecommand \bibfield  [0]{\@secondoftwo}%
\providecommand \translation [1]{[#1]}%
\providecommand \BibitemOpen [0]{}%
\providecommand \bibitemStop [0]{}%
\providecommand \bibitemNoStop [0]{.\EOS\space}%
\providecommand \EOS [0]{\spacefactor3000\relax}%
\providecommand \BibitemShut  [1]{\csname bibitem#1\endcsname}%
\let\auto@bib@innerbib\@empty
\bibitem [{\citenamefont {Lu}\ \emph {et~al.}(2014)\citenamefont {Lu},
  \citenamefont {Joannopoulos},\ and\ \citenamefont
  {Soljaclc}}]{Lu:2014-821:NPHOT}%
  \BibitemOpen
  \bibfield  {author} {\bibinfo {author} {\bibfnamefont {L.}~\bibnamefont
  {Lu}}, \bibinfo {author} {\bibfnamefont {J.~D.}\ \bibnamefont
  {Joannopoulos}}, \ and\ \bibinfo {author} {\bibfnamefont {M.}~\bibnamefont
  {Soljaclc}},\ }\bibfield  {title} {{\selectlanguage {English}\enquote
  {\bibinfo {title} {Topological photonics},}\ }}\href
  {https://doi.org/10.1038/NPHOTON.2014.248} {\bibfield  {journal} {\bibinfo
  {journal} {Nat. Photon.}\ }\textbf {\bibinfo {volume} {8}},\ \bibinfo {pages}
  {821--829} (\bibinfo {year} {2014})}\BibitemShut {NoStop}%
\bibitem [{\citenamefont {Christodoulides}\ \emph {et~al.}(2003)\citenamefont
  {Christodoulides}, \citenamefont {Lederer},\ and\ \citenamefont
  {Silberberg}}]{Christodoulides:2003-817:NAT}%
  \BibitemOpen
  \bibfield  {author} {\bibinfo {author} {\bibfnamefont {D.~N.}\ \bibnamefont
  {Christodoulides}}, \bibinfo {author} {\bibfnamefont {F.}~\bibnamefont
  {Lederer}}, \ and\ \bibinfo {author} {\bibfnamefont {Y.}~\bibnamefont
  {Silberberg}},\ }\bibfield  {title} {{\selectlanguage {English}\enquote
  {\bibinfo {title} {Discretizing light behaviour in linear and nonlinear
  waveguide lattices},}\ }}\href {https://doi.org/10.1038/nature01936}
  {\bibfield  {journal} {\bibinfo  {journal} {Nature}\ }\textbf {\bibinfo
  {volume} {424}},\ \bibinfo {pages} {817--823} (\bibinfo {year}
  {2003})}\BibitemShut {NoStop}%
\bibitem [{\citenamefont {Engheta}\ and\ \citenamefont
  {Ziolkowski}(2006)}]{Engheta:2006:Metamaterials}%
  \BibitemOpen
  \bibinfo {editor} {\bibfnamefont {N.}~\bibnamefont {Engheta}}\ and\ \bibinfo
  {editor} {\bibfnamefont {R.~W.}\ \bibnamefont {Ziolkowski}},\ eds.,\ \href
  {https://doi.org/10.1002/0471784192} {\emph {\bibinfo {title}
  {{Metamaterials}}}}\ (\bibinfo  {publisher} {John Wiley {\&} Sons, Inc.},\
  \bibinfo {address} {Hoboken, NJ, USA},\ \bibinfo {year} {2006})\BibitemShut
  {NoStop}%
\bibitem [{\citenamefont {Mittal}\ \emph {et~al.}(2014)\citenamefont {Mittal},
  \citenamefont {Fan}, \citenamefont {Faez}, \citenamefont {Migdall},
  \citenamefont {Taylor},\ and\ \citenamefont
  {Hafezi}}]{Mittal:2014-87403:PRL}%
  \BibitemOpen
  \bibfield  {author} {\bibinfo {author} {\bibfnamefont {S.}~\bibnamefont
  {Mittal}}, \bibinfo {author} {\bibfnamefont {J.}~\bibnamefont {Fan}},
  \bibinfo {author} {\bibfnamefont {S.}~\bibnamefont {Faez}}, \bibinfo {author}
  {\bibfnamefont {A.}~\bibnamefont {Migdall}}, \bibinfo {author} {\bibfnamefont
  {J.~M.}\ \bibnamefont {Taylor}}, \ and\ \bibinfo {author} {\bibfnamefont
  {M.}~\bibnamefont {Hafezi}},\ }\bibfield  {title} {{\selectlanguage
  {English}\enquote {\bibinfo {title} {Topologically robust transport of
  photons in a synthetic gauge field},}\ }}\href
  {https://doi.org/10.1103/PhysRevLett.113.087403} {\bibfield  {journal}
  {\bibinfo  {journal} {Phys. Rev. Lett.}\ }\textbf {\bibinfo {volume} {113}},\
  \bibinfo {pages} {087403} (\bibinfo {year} {2014})}\BibitemShut {NoStop}%
\bibitem [{\citenamefont {Yuan}\ \emph
  {et~al.}(2018{\natexlab{a}})\citenamefont {Yuan}, \citenamefont {Lin},
  \citenamefont {Xiao},\ and\ \citenamefont {Fan}}]{Yuan:2018-1396:OPT}%
  \BibitemOpen
  \bibfield  {author} {\bibinfo {author} {\bibfnamefont {L.~Q.}\ \bibnamefont
  {Yuan}}, \bibinfo {author} {\bibfnamefont {Q.}~\bibnamefont {Lin}}, \bibinfo
  {author} {\bibfnamefont {M.}~\bibnamefont {Xiao}}, \ and\ \bibinfo {author}
  {\bibfnamefont {S.~H.}\ \bibnamefont {Fan}},\ }\bibfield  {title}
  {{\selectlanguage {English}\enquote {\bibinfo {title} {Synthetic dimension in
  photonics},}\ }}\href {https://doi.org/10.1364/OPTICA.5.001396} {\bibfield
  {journal} {\bibinfo  {journal} {Optica}\ }\textbf {\bibinfo {volume} {5}},\
  \bibinfo {pages} {1396--1405} (\bibinfo {year}
  {2018}{\natexlab{a}})}\BibitemShut {NoStop}%
\bibitem [{\citenamefont {Ozawa}\ and\ \citenamefont
  {Price}(2019)}]{Ozawa:2019-349:NRP}%
  \BibitemOpen
  \bibfield  {author} {\bibinfo {author} {\bibfnamefont {T.}~\bibnamefont
  {Ozawa}}\ and\ \bibinfo {author} {\bibfnamefont {H.~M.}\ \bibnamefont
  {Price}},\ }\bibfield  {title} {\enquote {\bibinfo {title} {{Topological
  quantum matter in synthetic dimensions}},}\ }\href
  {https://doi.org/10.1038/s42254-019-0045-3} {\bibfield  {journal} {\bibinfo
  {journal} {Nat. Rev. Phys.}\ }\textbf {\bibinfo {volume} {1}},\ \bibinfo
  {pages} {349--357} (\bibinfo {year} {2019})}\BibitemShut {NoStop}%
\bibitem [{\citenamefont {Boada}\ \emph {et~al.}(2012)\citenamefont {Boada},
  \citenamefont {Celi}, \citenamefont {Latorre},\ and\ \citenamefont
  {Lewenstein}}]{Boada:2012-133001:PRL}%
  \BibitemOpen
  \bibfield  {author} {\bibinfo {author} {\bibfnamefont {O.}~\bibnamefont
  {Boada}}, \bibinfo {author} {\bibfnamefont {A.}~\bibnamefont {Celi}},
  \bibinfo {author} {\bibfnamefont {J.~I.}\ \bibnamefont {Latorre}}, \ and\
  \bibinfo {author} {\bibfnamefont {M.}~\bibnamefont {Lewenstein}},\ }\bibfield
   {title} {{\selectlanguage {English}\enquote {\bibinfo {title} {Quantum
  simulation of an extra dimension},}\ }}\href
  {https://doi.org/10.1103/PhysRevLett.108.133001} {\bibfield  {journal}
  {\bibinfo  {journal} {Phys. Rev. Lett.}\ }\textbf {\bibinfo {volume} {108}},\
  \bibinfo {pages} {133001} (\bibinfo {year} {2012})}\BibitemShut {NoStop}%
\bibitem [{\citenamefont {Celi}\ \emph {et~al.}(2014)\citenamefont {Celi},
  \citenamefont {Massignan}, \citenamefont {Ruseckas}, \citenamefont {Goldman},
  \citenamefont {Spielman}, \citenamefont {Juzeliunas},\ and\ \citenamefont
  {Lewenstein}}]{Celi:2014-43001:PRL}%
  \BibitemOpen
  \bibfield  {author} {\bibinfo {author} {\bibfnamefont {A.}~\bibnamefont
  {Celi}}, \bibinfo {author} {\bibfnamefont {P.}~\bibnamefont {Massignan}},
  \bibinfo {author} {\bibfnamefont {J.}~\bibnamefont {Ruseckas}}, \bibinfo
  {author} {\bibfnamefont {N.}~\bibnamefont {Goldman}}, \bibinfo {author}
  {\bibfnamefont {I.~B.}\ \bibnamefont {Spielman}}, \bibinfo {author}
  {\bibfnamefont {G.}~\bibnamefont {Juzeliunas}}, \ and\ \bibinfo {author}
  {\bibfnamefont {M.}~\bibnamefont {Lewenstein}},\ }\bibfield  {title}
  {{\selectlanguage {English}\enquote {\bibinfo {title} {Synthetic gauge fields
  in synthetic dimensions},}\ }}\href
  {https://doi.org/10.1103/PhysRevLett.112.043001} {\bibfield  {journal}
  {\bibinfo  {journal} {Phys. Rev. Lett.}\ }\textbf {\bibinfo {volume} {112}},\
  \bibinfo {pages} {043001} (\bibinfo {year} {2014})}\BibitemShut {NoStop}%
\bibitem [{\citenamefont {Yuan}\ \emph {et~al.}(2016)\citenamefont {Yuan},
  \citenamefont {Shi},\ and\ \citenamefont {Fan}}]{Yuan:2016-741:OL}%
  \BibitemOpen
  \bibfield  {author} {\bibinfo {author} {\bibfnamefont {L.~Q.}\ \bibnamefont
  {Yuan}}, \bibinfo {author} {\bibfnamefont {Y.}~\bibnamefont {Shi}}, \ and\
  \bibinfo {author} {\bibfnamefont {S.~H.}\ \bibnamefont {Fan}},\ }\bibfield
  {title} {{\selectlanguage {English}\enquote {\bibinfo {title} {Photonic gauge
  potential in a system with a synthetic frequency dimension},}\ }}\href
  {https://doi.org/10.1364/OL.41.000741} {\bibfield  {journal} {\bibinfo
  {journal} {Opt. Lett.}\ }\textbf {\bibinfo {volume} {41}},\ \bibinfo {pages}
  {741--744} (\bibinfo {year} {2016})}\BibitemShut {NoStop}%
\bibitem [{\citenamefont {Ozawa}\ \emph {et~al.}(2016)\citenamefont {Ozawa},
  \citenamefont {Price}, \citenamefont {Goldman}, \citenamefont {Zilberberg},\
  and\ \citenamefont {Carusotto}}]{Ozawa:2016-43827:PRA}%
  \BibitemOpen
  \bibfield  {author} {\bibinfo {author} {\bibfnamefont {T.}~\bibnamefont
  {Ozawa}}, \bibinfo {author} {\bibfnamefont {H.~M.}\ \bibnamefont {Price}},
  \bibinfo {author} {\bibfnamefont {N.}~\bibnamefont {Goldman}}, \bibinfo
  {author} {\bibfnamefont {O.}~\bibnamefont {Zilberberg}}, \ and\ \bibinfo
  {author} {\bibfnamefont {I.}~\bibnamefont {Carusotto}},\ }\bibfield  {title}
  {{\selectlanguage {English}\enquote {\bibinfo {title} {Synthetic dimensions
  in integrated photonics: From optical isolation to four-dimensional quantum
  {H}all physics},}\ }}\href {https://doi.org/10.1103/PhysRevA.93.043827}
  {\bibfield  {journal} {\bibinfo  {journal} {Phys. Rev. A}\ }\textbf {\bibinfo
  {volume} {93}},\ \bibinfo {pages} {043827} (\bibinfo {year}
  {2016})}\BibitemShut {NoStop}%
\bibitem [{\citenamefont {Martin}\ \emph {et~al.}(2017)\citenamefont {Martin},
  \citenamefont {Refael},\ and\ \citenamefont
  {Halperin}}]{Martin:2017-41008:PRX}%
  \BibitemOpen
  \bibfield  {author} {\bibinfo {author} {\bibfnamefont {I.}~\bibnamefont
  {Martin}}, \bibinfo {author} {\bibfnamefont {G.}~\bibnamefont {Refael}}, \
  and\ \bibinfo {author} {\bibfnamefont {B.}~\bibnamefont {Halperin}},\
  }\bibfield  {title} {{\selectlanguage {English}\enquote {\bibinfo {title}
  {Topological frequency conversion in strongly driven quantum systems},}\
  }}\href {https://doi.org/10.1103/PhysRevX.7.041008} {\bibfield  {journal}
  {\bibinfo  {journal} {Phys. Rev. X}\ }\textbf {\bibinfo {volume} {7}},\
  \bibinfo {pages} {041008} (\bibinfo {year} {2017})}\BibitemShut {NoStop}%
\bibitem [{\citenamefont {Regensburger}\ \emph {et~al.}(2012)\citenamefont
  {Regensburger}, \citenamefont {Bersch}, \citenamefont {Miri}, \citenamefont
  {Onishchukov}, \citenamefont {Christodoulides},\ and\ \citenamefont
  {Peschel}}]{Regensburger:2012-167:NAT}%
  \BibitemOpen
  \bibfield  {author} {\bibinfo {author} {\bibfnamefont {A.}~\bibnamefont
  {Regensburger}}, \bibinfo {author} {\bibfnamefont {C.}~\bibnamefont
  {Bersch}}, \bibinfo {author} {\bibfnamefont {M.~A.}\ \bibnamefont {Miri}},
  \bibinfo {author} {\bibfnamefont {G.}~\bibnamefont {Onishchukov}}, \bibinfo
  {author} {\bibfnamefont {D.~N.}\ \bibnamefont {Christodoulides}}, \ and\
  \bibinfo {author} {\bibfnamefont {U.}~\bibnamefont {Peschel}},\ }\bibfield
  {title} {{\selectlanguage {English}\enquote {\bibinfo {title} {Parity-time
  synthetic photonic lattices},}\ }}\href {https://doi.org/10.1038/nature11298}
  {\bibfield  {journal} {\bibinfo  {journal} {Nature}\ }\textbf {\bibinfo
  {volume} {488}},\ \bibinfo {pages} {167--171} (\bibinfo {year}
  {2012})}\BibitemShut {NoStop}%
\bibitem [{\citenamefont {Stuhl}\ \emph {et~al.}(2015)\citenamefont {Stuhl},
  \citenamefont {Lu}, \citenamefont {Aycock}, \citenamefont {Genkina},\ and\
  \citenamefont {Spielman}}]{Stuhl:2015-1514:SCI}%
  \BibitemOpen
  \bibfield  {author} {\bibinfo {author} {\bibfnamefont {B.~K.}\ \bibnamefont
  {Stuhl}}, \bibinfo {author} {\bibfnamefont {H.~I.}\ \bibnamefont {Lu}},
  \bibinfo {author} {\bibfnamefont {L.~M.}\ \bibnamefont {Aycock}}, \bibinfo
  {author} {\bibfnamefont {D.}~\bibnamefont {Genkina}}, \ and\ \bibinfo
  {author} {\bibfnamefont {I.~B.}\ \bibnamefont {Spielman}},\ }\bibfield
  {title} {{\selectlanguage {English}\enquote {\bibinfo {title} {Visualizing
  edge states with an atomic {B}ose gas in the quantum {H}all regime},}\
  }}\href {https://doi.org/10.1126/science.aaa8515} {\bibfield  {journal}
  {\bibinfo  {journal} {Science}\ }\textbf {\bibinfo {volume} {349}},\ \bibinfo
  {pages} {1514--1517} (\bibinfo {year} {2015})}\BibitemShut {NoStop}%
\bibitem [{\citenamefont {Mancini}\ \emph {et~al.}(2015)\citenamefont
  {Mancini}, \citenamefont {Pagano}, \citenamefont {Cappellini}, \citenamefont
  {Livi}, \citenamefont {Rider}, \citenamefont {Catani}, \citenamefont {Sias},
  \citenamefont {Zoller}, \citenamefont {Inguscio}, \citenamefont {Dalmonte},\
  and\ \citenamefont {Fallani}}]{Mancini:2015-1510:SCI}%
  \BibitemOpen
  \bibfield  {author} {\bibinfo {author} {\bibfnamefont {M.}~\bibnamefont
  {Mancini}}, \bibinfo {author} {\bibfnamefont {G.}~\bibnamefont {Pagano}},
  \bibinfo {author} {\bibfnamefont {G.}~\bibnamefont {Cappellini}}, \bibinfo
  {author} {\bibfnamefont {L.}~\bibnamefont {Livi}}, \bibinfo {author}
  {\bibfnamefont {M.}~\bibnamefont {Rider}}, \bibinfo {author} {\bibfnamefont
  {J.}~\bibnamefont {Catani}}, \bibinfo {author} {\bibfnamefont
  {C.}~\bibnamefont {Sias}}, \bibinfo {author} {\bibfnamefont {P.}~\bibnamefont
  {Zoller}}, \bibinfo {author} {\bibfnamefont {M.}~\bibnamefont {Inguscio}},
  \bibinfo {author} {\bibfnamefont {M.}~\bibnamefont {Dalmonte}}, \ and\
  \bibinfo {author} {\bibfnamefont {L.}~\bibnamefont {Fallani}},\ }\bibfield
  {title} {{\selectlanguage {English}\enquote {\bibinfo {title} {Observation of
  chiral edge states with neutral fermions in synthetic {H}all ribbons},}\
  }}\href {https://doi.org/10.1126/science.aaa8736} {\bibfield  {journal}
  {\bibinfo  {journal} {Science}\ }\textbf {\bibinfo {volume} {349}},\ \bibinfo
  {pages} {1510--1513} (\bibinfo {year} {2015})}\BibitemShut {NoStop}%
\bibitem [{\citenamefont {Lustig}\ \emph {et~al.}(2019)\citenamefont {Lustig},
  \citenamefont {Weimann}, \citenamefont {Plotnik}, \citenamefont {Lumer},
  \citenamefont {Bandres}, \citenamefont {Szameit},\ and\ \citenamefont
  {Segev}}]{Lustig:2019-356:NAT}%
  \BibitemOpen
  \bibfield  {author} {\bibinfo {author} {\bibfnamefont {E.}~\bibnamefont
  {Lustig}}, \bibinfo {author} {\bibfnamefont {S.}~\bibnamefont {Weimann}},
  \bibinfo {author} {\bibfnamefont {Y.}~\bibnamefont {Plotnik}}, \bibinfo
  {author} {\bibfnamefont {Y.}~\bibnamefont {Lumer}}, \bibinfo {author}
  {\bibfnamefont {M.~A.}\ \bibnamefont {Bandres}}, \bibinfo {author}
  {\bibfnamefont {A.}~\bibnamefont {Szameit}}, \ and\ \bibinfo {author}
  {\bibfnamefont {M.}~\bibnamefont {Segev}},\ }\bibfield  {title}
  {{\selectlanguage {English}\enquote {\bibinfo {title} {Photonic topological
  insulator in synthetic dimensions},}\ }}\href
  {https://doi.org/10.1038/s41586-019-0943-7} {\bibfield  {journal} {\bibinfo
  {journal} {Nature}\ }\textbf {\bibinfo {volume} {567}},\ \bibinfo {pages}
  {356--371} (\bibinfo {year} {2019})}\BibitemShut {NoStop}%
\bibitem [{\citenamefont {Maczewsky}\ \emph {et~al.}(2020)\citenamefont
  {Maczewsky}, \citenamefont {Wang}, \citenamefont {Dovgiy}, \citenamefont
  {Miroshnichenko}, \citenamefont {Moroz}, \citenamefont {Ehrhardt},
  \citenamefont {Heinrich}, \citenamefont {Christodoulides}, \citenamefont
  {Szameit},\ and\ \citenamefont {Sukhorukov}}]{Maczewsky:2020-76:NPHOT}%
  \BibitemOpen
  \bibfield  {author} {\bibinfo {author} {\bibfnamefont {L.~J.}\ \bibnamefont
  {Maczewsky}}, \bibinfo {author} {\bibfnamefont {K.}~\bibnamefont {Wang}},
  \bibinfo {author} {\bibfnamefont {A.~A.}\ \bibnamefont {Dovgiy}}, \bibinfo
  {author} {\bibfnamefont {A.~E.}\ \bibnamefont {Miroshnichenko}}, \bibinfo
  {author} {\bibfnamefont {A.}~\bibnamefont {Moroz}}, \bibinfo {author}
  {\bibfnamefont {M.}~\bibnamefont {Ehrhardt}}, \bibinfo {author}
  {\bibfnamefont {M.}~\bibnamefont {Heinrich}}, \bibinfo {author}
  {\bibfnamefont {D.~N.}\ \bibnamefont {Christodoulides}}, \bibinfo {author}
  {\bibfnamefont {A.}~\bibnamefont {Szameit}}, \ and\ \bibinfo {author}
  {\bibfnamefont {A.~A.}\ \bibnamefont {Sukhorukov}},\ }\bibfield  {title}
  {\enquote {\bibinfo {title} {{Synthesizing multi-dimensional excitation
  dynamics and localization transition in one-dimensional lattices}},}\ }\href
  {https://doi.org/10.1038/s41566-019-0562-8} {\bibfield  {journal} {\bibinfo
  {journal} {Nat. Photon.}\ }\textbf {\bibinfo {volume} {14}},\ \bibinfo
  {pages} {76--81} (\bibinfo {year} {2020})}\BibitemShut {NoStop}%
\bibitem [{\citenamefont {Casanova}\ \emph {et~al.}(2012)\citenamefont
  {Casanova}, \citenamefont {Mezzacapo}, \citenamefont {Lamata},\ and\
  \citenamefont {Solano}}]{Casanova:2012-190502:PRL}%
  \BibitemOpen
  \bibfield  {author} {\bibinfo {author} {\bibfnamefont {J.}~\bibnamefont
  {Casanova}}, \bibinfo {author} {\bibfnamefont {A.}~\bibnamefont {Mezzacapo}},
  \bibinfo {author} {\bibfnamefont {L.}~\bibnamefont {Lamata}}, \ and\ \bibinfo
  {author} {\bibfnamefont {E.}~\bibnamefont {Solano}},\ }\bibfield  {title}
  {{\selectlanguage {English}\enquote {\bibinfo {title} {Quantum simulation of
  interacting fermion lattice models in trapped ions},}\ }}\href
  {https://doi.org/10.1103/PhysRevLett.108.190502} {\bibfield  {journal}
  {\bibinfo  {journal} {Phys. Rev. Lett.}\ }\textbf {\bibinfo {volume} {108}},\
  \bibinfo {pages} {190502} (\bibinfo {year} {2012})}\BibitemShut {NoStop}%
\bibitem [{\citenamefont {Grass}\ \emph {et~al.}(2015)\citenamefont {Grass},
  \citenamefont {Muschik}, \citenamefont {Celi}, \citenamefont {Chhajlany},\
  and\ \citenamefont {Lewenstein}}]{Grass:2015-63612:PRA}%
  \BibitemOpen
  \bibfield  {author} {\bibinfo {author} {\bibfnamefont {T.}~\bibnamefont
  {Grass}}, \bibinfo {author} {\bibfnamefont {C.}~\bibnamefont {Muschik}},
  \bibinfo {author} {\bibfnamefont {A.}~\bibnamefont {Celi}}, \bibinfo {author}
  {\bibfnamefont {R.~W.}\ \bibnamefont {Chhajlany}}, \ and\ \bibinfo {author}
  {\bibfnamefont {M.}~\bibnamefont {Lewenstein}},\ }\bibfield  {title}
  {{\selectlanguage {English}\enquote {\bibinfo {title} {Synthetic magnetic
  fluxes and topological order in one-dimensional spin systems},}\ }}\href
  {https://doi.org/10.1103/PhysRevA.91.063612} {\bibfield  {journal} {\bibinfo
  {journal} {Phys. Rev. A}\ }\textbf {\bibinfo {volume} {91}},\ \bibinfo
  {pages} {063612} (\bibinfo {year} {2015})}\BibitemShut {NoStop}%
\bibitem [{\citenamefont {Yuan}\ \emph
  {et~al.}(2018{\natexlab{b}})\citenamefont {Yuan}, \citenamefont {Xiao},
  \citenamefont {Lin},\ and\ \citenamefont {Fan}}]{Yuan:2018-104105:PRB}%
  \BibitemOpen
  \bibfield  {author} {\bibinfo {author} {\bibfnamefont {L.~Q.}\ \bibnamefont
  {Yuan}}, \bibinfo {author} {\bibfnamefont {M.}~\bibnamefont {Xiao}}, \bibinfo
  {author} {\bibfnamefont {Q.}~\bibnamefont {Lin}}, \ and\ \bibinfo {author}
  {\bibfnamefont {S.~H.}\ \bibnamefont {Fan}},\ }\bibfield  {title}
  {{\selectlanguage {English}\enquote {\bibinfo {title} {Synthetic space with
  arbitrary dimensions in a few rings undergoing dynamic modulation},}\ }}\href
  {https://doi.org/10.1103/PhysRevB.97.104105} {\bibfield  {journal} {\bibinfo
  {journal} {Phys. Rev. B}\ }\textbf {\bibinfo {volume} {97}},\ \bibinfo
  {pages} {104105} (\bibinfo {year} {2018}{\natexlab{b}})}\BibitemShut
  {NoStop}%
\bibitem [{\citenamefont {Dutt}\ \emph
  {et~al.}(2019{\natexlab{a}})\citenamefont {Dutt}, \citenamefont {Minkov},\
  and\ \citenamefont {Fan}}]{Dutt:1911.11310:ARXIV}%
  \BibitemOpen
  \bibfield  {author} {\bibinfo {author} {\bibfnamefont {A.}~\bibnamefont
  {Dutt}}, \bibinfo {author} {\bibfnamefont {M.}~\bibnamefont {Minkov}}, \ and\
  \bibinfo {author} {\bibfnamefont {S.}~\bibnamefont {Fan}},\ }\bibfield
  {title} {\enquote {\bibinfo {title} {{Higher-order topological insulators in
  synthetic dimensions}},}\ }\href {http://arxiv.org/abs/1911.11310} {\bibfield
   {journal} {\bibinfo  {journal} {arXiv}\ }\textbf {\bibinfo {volume}
  {\mdseries 1911.11310}} (\bibinfo {year} {2019}{\natexlab{a}})}\BibitemShut
  {NoStop}%
\bibitem [{\citenamefont {Artyukhov}\ \emph {et~al.}(2014)\citenamefont
  {Artyukhov}, \citenamefont {Penev},\ and\ \citenamefont
  {Yakobson}}]{Artyukhov:2014-4892:NCOM}%
  \BibitemOpen
  \bibfield  {author} {\bibinfo {author} {\bibfnamefont {V.~I.}\ \bibnamefont
  {Artyukhov}}, \bibinfo {author} {\bibfnamefont {E.~S.}\ \bibnamefont
  {Penev}}, \ and\ \bibinfo {author} {\bibfnamefont {B.~I.}\ \bibnamefont
  {Yakobson}},\ }\bibfield  {title} {{\selectlanguage {English}\enquote
  {\bibinfo {title} {Why nanotubes grow chiral},}\ }}\href
  {https://doi.org/10.1038/ncomms5892} {\bibfield  {journal} {\bibinfo
  {journal} {Nat. Commun.}\ }\textbf {\bibinfo {volume} {5}},\ \bibinfo {pages}
  {4892} (\bibinfo {year} {2014})}\BibitemShut {NoStop}%
\bibitem [{\citenamefont {Bell}\ \emph {et~al.}(2017)\citenamefont {Bell},
  \citenamefont {Wang}, \citenamefont {Solntsev}, \citenamefont {Neshev},
  \citenamefont {Sukhorukov},\ and\ \citenamefont
  {Eggleton}}]{Bell:2017-1433:OPT}%
  \BibitemOpen
  \bibfield  {author} {\bibinfo {author} {\bibfnamefont {B.~A.}\ \bibnamefont
  {Bell}}, \bibinfo {author} {\bibfnamefont {K.}~\bibnamefont {Wang}}, \bibinfo
  {author} {\bibfnamefont {A.~S.}\ \bibnamefont {Solntsev}}, \bibinfo {author}
  {\bibfnamefont {D.~N.}\ \bibnamefont {Neshev}}, \bibinfo {author}
  {\bibfnamefont {A.~A.}\ \bibnamefont {Sukhorukov}}, \ and\ \bibinfo {author}
  {\bibfnamefont {B.~J.}\ \bibnamefont {Eggleton}},\ }\bibfield  {title}
  {{\selectlanguage {English}\enquote {\bibinfo {title} {Spectral photonic
  lattices with complex long-range coupling},}\ }}\href
  {https://doi.org/10.1364/OPTICA.4.001433} {\bibfield  {journal} {\bibinfo
  {journal} {Optica}\ }\textbf {\bibinfo {volume} {4}},\ \bibinfo {pages}
  {1433--1436} (\bibinfo {year} {2017})}\BibitemShut {NoStop}%
\bibitem [{\citenamefont {Qin}\ \emph {et~al.}(2018)\citenamefont {Qin},
  \citenamefont {Zhou}, \citenamefont {Peng}, \citenamefont {Sounas},
  \citenamefont {Zhu}, \citenamefont {Wang}, \citenamefont {Dong},
  \citenamefont {Zhang}, \citenamefont {Alu},\ and\ \citenamefont
  {Lu}}]{Qin:2018-133901:PRL}%
  \BibitemOpen
  \bibfield  {author} {\bibinfo {author} {\bibfnamefont {C.~Z.}\ \bibnamefont
  {Qin}}, \bibinfo {author} {\bibfnamefont {F.}~\bibnamefont {Zhou}}, \bibinfo
  {author} {\bibfnamefont {Y.~G.}\ \bibnamefont {Peng}}, \bibinfo {author}
  {\bibfnamefont {D.}~\bibnamefont {Sounas}}, \bibinfo {author} {\bibfnamefont
  {X.~F.}\ \bibnamefont {Zhu}}, \bibinfo {author} {\bibfnamefont
  {B.}~\bibnamefont {Wang}}, \bibinfo {author} {\bibfnamefont {J.~J.}\
  \bibnamefont {Dong}}, \bibinfo {author} {\bibfnamefont {X.~L.}\ \bibnamefont
  {Zhang}}, \bibinfo {author} {\bibfnamefont {A.}~\bibnamefont {Alu}}, \ and\
  \bibinfo {author} {\bibfnamefont {P.~X.}\ \bibnamefont {Lu}},\ }\bibfield
  {title} {{\selectlanguage {English}\enquote {\bibinfo {title} {Spectrum
  control through discrete frequency diffraction in the presence of photonic
  gauge potentials},}\ }}\href {https://doi.org/10.1103/PhysRevLett.120.133901}
  {\bibfield  {journal} {\bibinfo  {journal} {Phys. Rev. Lett.}\ }\textbf
  {\bibinfo {volume} {120}},\ \bibinfo {pages} {133901} (\bibinfo {year}
  {2018})}\BibitemShut {NoStop}%
\bibitem [{\citenamefont {Lumer}\ \emph {et~al.}(2019)\citenamefont {Lumer},
  \citenamefont {Bandres}, \citenamefont {Heinrich}, \citenamefont {Maczewsky},
  \citenamefont {Herzig-Sheinfux}, \citenamefont {Szameit},\ and\ \citenamefont
  {Segev}}]{Lumer:2019-339:NPHOT}%
  \BibitemOpen
  \bibfield  {author} {\bibinfo {author} {\bibfnamefont {Y.}~\bibnamefont
  {Lumer}}, \bibinfo {author} {\bibfnamefont {M.~A.}\ \bibnamefont {Bandres}},
  \bibinfo {author} {\bibfnamefont {M.}~\bibnamefont {Heinrich}}, \bibinfo
  {author} {\bibfnamefont {L.~J.}\ \bibnamefont {Maczewsky}}, \bibinfo {author}
  {\bibfnamefont {H.}~\bibnamefont {Herzig-Sheinfux}}, \bibinfo {author}
  {\bibfnamefont {A.}~\bibnamefont {Szameit}}, \ and\ \bibinfo {author}
  {\bibfnamefont {M.}~\bibnamefont {Segev}},\ }\bibfield  {title}
  {{\selectlanguage {English}\enquote {\bibinfo {title} {Light guiding by
  artificial gauge fields},}\ }}\href
  {https://doi.org/10.1038/s41566-019-0370-1} {\bibfield  {journal} {\bibinfo
  {journal} {Nat. Photon.}\ }\textbf {\bibinfo {volume} {13}},\ \bibinfo
  {pages} {339--346} (\bibinfo {year} {2019})}\BibitemShut {NoStop}%
\bibitem [{\citenamefont {Lin}\ and\ \citenamefont
  {Fan}(2014)}]{Lin:2014-31031:PRX}%
  \BibitemOpen
  \bibfield  {author} {\bibinfo {author} {\bibfnamefont {Q.}~\bibnamefont
  {Lin}}\ and\ \bibinfo {author} {\bibfnamefont {S.~H.}\ \bibnamefont {Fan}},\
  }\bibfield  {title} {{\selectlanguage {English}\enquote {\bibinfo {title}
  {Light guiding by effective gauge field for photons},}\ }}\href
  {https://doi.org/10.1103/PhysRevX.4.031031} {\bibfield  {journal} {\bibinfo
  {journal} {Phys. Rev. X}\ }\textbf {\bibinfo {volume} {4}},\ \bibinfo {pages}
  {031031} (\bibinfo {year} {2014})}\BibitemShut {NoStop}%
\bibitem [{\citenamefont {Harari}\ \emph {et~al.}(2018)\citenamefont {Harari},
  \citenamefont {Bandres}, \citenamefont {Lumer}, \citenamefont {Rechtsman},
  \citenamefont {Chong}, \citenamefont {Khajavikhan}, \citenamefont
  {Christodoulides},\ and\ \citenamefont {Segev}}]{Harari:2018-eaar4003:SCI}%
  \BibitemOpen
  \bibfield  {author} {\bibinfo {author} {\bibfnamefont {G.}~\bibnamefont
  {Harari}}, \bibinfo {author} {\bibfnamefont {M.~A.}\ \bibnamefont {Bandres}},
  \bibinfo {author} {\bibfnamefont {Y.}~\bibnamefont {Lumer}}, \bibinfo
  {author} {\bibfnamefont {M.~C.}\ \bibnamefont {Rechtsman}}, \bibinfo {author}
  {\bibfnamefont {Y.~D.}\ \bibnamefont {Chong}}, \bibinfo {author}
  {\bibfnamefont {M.}~\bibnamefont {Khajavikhan}}, \bibinfo {author}
  {\bibfnamefont {D.~N.}\ \bibnamefont {Christodoulides}}, \ and\ \bibinfo
  {author} {\bibfnamefont {M.}~\bibnamefont {Segev}},\ }\bibfield  {title}
  {{\selectlanguage {English}\enquote {\bibinfo {title} {Topological insulator
  laser: Theory},}\ }}\href {https://doi.org/10.1126/science.aar4003}
  {\bibfield  {journal} {\bibinfo  {journal} {Science}\ }\textbf {\bibinfo
  {volume} {359}},\ \bibinfo {pages} {eaar4003} (\bibinfo {year}
  {2018})}\BibitemShut {NoStop}%
\bibitem [{\citenamefont {Bandres}\ \emph {et~al.}(2018)\citenamefont
  {Bandres}, \citenamefont {Wittek}, \citenamefont {Harari}, \citenamefont
  {Parto}, \citenamefont {Ren}, \citenamefont {Segev}, \citenamefont
  {Christodoulides},\ and\ \citenamefont
  {Khajavilchan}}]{Bandres:2018-eaar4005:SCI}%
  \BibitemOpen
  \bibfield  {author} {\bibinfo {author} {\bibfnamefont {M.~A.}\ \bibnamefont
  {Bandres}}, \bibinfo {author} {\bibfnamefont {S.}~\bibnamefont {Wittek}},
  \bibinfo {author} {\bibfnamefont {G.}~\bibnamefont {Harari}}, \bibinfo
  {author} {\bibfnamefont {M.}~\bibnamefont {Parto}}, \bibinfo {author}
  {\bibfnamefont {J.~H.}\ \bibnamefont {Ren}}, \bibinfo {author} {\bibfnamefont
  {M.}~\bibnamefont {Segev}}, \bibinfo {author} {\bibfnamefont {D.~N.}\
  \bibnamefont {Christodoulides}}, \ and\ \bibinfo {author} {\bibfnamefont
  {M.}~\bibnamefont {Khajavilchan}},\ }\bibfield  {title} {{\selectlanguage
  {English}\enquote {\bibinfo {title} {Topological insulator laser:
  Experiments},}\ }}\href {https://doi.org/10.1126/science.aar4005} {\bibfield
  {journal} {\bibinfo  {journal} {Science}\ }\textbf {\bibinfo {volume}
  {359}},\ \bibinfo {pages} {eaar4005} (\bibinfo {year} {2018})}\BibitemShut
  {NoStop}%
\bibitem [{\citenamefont {Dutt}\ \emph {et~al.}(2020)\citenamefont {Dutt},
  \citenamefont {Lin}, \citenamefont {Yuan}, \citenamefont {Minkov},
  \citenamefont {Xiao},\ and\ \citenamefont {Fan}}]{Dutt:2020-59:SCI}%
  \BibitemOpen
  \bibfield  {author} {\bibinfo {author} {\bibfnamefont {A.}~\bibnamefont
  {Dutt}}, \bibinfo {author} {\bibfnamefont {Q.}~\bibnamefont {Lin}}, \bibinfo
  {author} {\bibfnamefont {L.~Q.}\ \bibnamefont {Yuan}}, \bibinfo {author}
  {\bibfnamefont {M.}~\bibnamefont {Minkov}}, \bibinfo {author} {\bibfnamefont
  {M.}~\bibnamefont {Xiao}}, \ and\ \bibinfo {author} {\bibfnamefont {S.~H.}\
  \bibnamefont {Fan}},\ }\bibfield  {title} {{\selectlanguage {English}\enquote
  {\bibinfo {title} {A single photonic cavity with two independent physical
  synthetic dimensions},}\ }}\href {https://doi.org/10.1126/science.aaz3071}
  {\bibfield  {journal} {\bibinfo  {journal} {Science}\ }\textbf {\bibinfo
  {volume} {367}},\ \bibinfo {pages} {59--89} (\bibinfo {year}
  {2020})}\BibitemShut {NoStop}%
\bibitem [{\citenamefont {Haldane}(1988)}]{Haldane:1988-2015:PRL}%
  \BibitemOpen
  \bibfield  {author} {\bibinfo {author} {\bibfnamefont {F.~D.~M.}\
  \bibnamefont {Haldane}},\ }\bibfield  {title} {{\selectlanguage
  {English}\enquote {\bibinfo {title} {Model for a quantum {H}all-effect
  without {L}andau-levels - condensed-matter realization of the parity
  anomaly},}\ }}\href {https://doi.org/10.1103/PhysRevLett.61.2015} {\bibfield
  {journal} {\bibinfo  {journal} {Phys. Rev. Lett.}\ }\textbf {\bibinfo
  {volume} {61}},\ \bibinfo {pages} {2015--2018} (\bibinfo {year}
  {1988})}\BibitemShut {NoStop}%
\bibitem [{\citenamefont {Razzari}\ \emph {et~al.}(2010)\citenamefont
  {Razzari}, \citenamefont {Duchesne}, \citenamefont {Ferrera}, \citenamefont
  {Morandotti}, \citenamefont {Chu}, \citenamefont {Little},\ and\
  \citenamefont {Moss}}]{Razzari:2010-41:NPHOT}%
  \BibitemOpen
  \bibfield  {author} {\bibinfo {author} {\bibfnamefont {L.}~\bibnamefont
  {Razzari}}, \bibinfo {author} {\bibfnamefont {D.}~\bibnamefont {Duchesne}},
  \bibinfo {author} {\bibfnamefont {M.}~\bibnamefont {Ferrera}}, \bibinfo
  {author} {\bibfnamefont {R.}~\bibnamefont {Morandotti}}, \bibinfo {author}
  {\bibfnamefont {S.}~\bibnamefont {Chu}}, \bibinfo {author} {\bibfnamefont
  {B.~E.}\ \bibnamefont {Little}}, \ and\ \bibinfo {author} {\bibfnamefont
  {D.~J.}\ \bibnamefont {Moss}},\ }\bibfield  {title} {{\selectlanguage
  {English}\enquote {\bibinfo {title} {Cmos-compatible integrated optical
  hyper-parametric oscillator},}\ }}\href
  {https://doi.org/10.1038/NPHOTON.2009.236} {\bibfield  {journal} {\bibinfo
  {journal} {Nat. Photon.}\ }\textbf {\bibinfo {volume} {4}},\ \bibinfo {pages}
  {41--45} (\bibinfo {year} {2010})}\BibitemShut {NoStop}%
\bibitem [{\citenamefont {Dutt}\ \emph
  {et~al.}(2019{\natexlab{b}})\citenamefont {Dutt}, \citenamefont {Minkov},
  \citenamefont {Lin}, \citenamefont {Yuan}, \citenamefont {Miller},\ and\
  \citenamefont {Fan}}]{Dutt:2019-162:ACSP}%
  \BibitemOpen
  \bibfield  {author} {\bibinfo {author} {\bibfnamefont {A.}~\bibnamefont
  {Dutt}}, \bibinfo {author} {\bibfnamefont {M.}~\bibnamefont {Minkov}},
  \bibinfo {author} {\bibfnamefont {Q.}~\bibnamefont {Lin}}, \bibinfo {author}
  {\bibfnamefont {L.~Q.}\ \bibnamefont {Yuan}}, \bibinfo {author}
  {\bibfnamefont {D.~A.~B.}\ \bibnamefont {Miller}}, \ and\ \bibinfo {author}
  {\bibfnamefont {S.~H.}\ \bibnamefont {Fan}},\ }\bibfield  {title}
  {{\selectlanguage {English}\enquote {\bibinfo {title} {Experimental
  demonstration of dynamical input isolation in nonadiabatically modulated
  photonic cavities},}\ }}\href {https://doi.org/10.1021/acsphotonics.8b01310}
  {\bibfield  {journal} {\bibinfo  {journal} {ACS Photonics}\ }\textbf
  {\bibinfo {volume} {6}},\ \bibinfo {pages} {162--169} (\bibinfo {year}
  {2019}{\natexlab{b}})}\BibitemShut {NoStop}%
\bibitem [{\citenamefont {Dutt}\ \emph
  {et~al.}(2019{\natexlab{c}})\citenamefont {Dutt}, \citenamefont {Minkov},
  \citenamefont {Lin}, \citenamefont {Yuan}, \citenamefont {Miller},\ and\
  \citenamefont {Fan}}]{Dutt:2019-3122:NCOM}%
  \BibitemOpen
  \bibfield  {author} {\bibinfo {author} {\bibfnamefont {A.}~\bibnamefont
  {Dutt}}, \bibinfo {author} {\bibfnamefont {M.}~\bibnamefont {Minkov}},
  \bibinfo {author} {\bibfnamefont {Q.}~\bibnamefont {Lin}}, \bibinfo {author}
  {\bibfnamefont {L.~Q.}\ \bibnamefont {Yuan}}, \bibinfo {author}
  {\bibfnamefont {D.~A.~B.}\ \bibnamefont {Miller}}, \ and\ \bibinfo {author}
  {\bibfnamefont {S.~H.}\ \bibnamefont {Fan}},\ }\bibfield  {title}
  {{\selectlanguage {English}\enquote {\bibinfo {title} {Experimental band
  structure spectroscopy along a synthetic dimension},}\ }}\href
  {https://doi.org/10.1038/s41467-019-11117-9} {\bibfield  {journal} {\bibinfo
  {journal} {Nat. Commun.}\ }\textbf {\bibinfo {volume} {10}},\ \bibinfo
  {pages} {3122} (\bibinfo {year} {2019}{\natexlab{c}})}\BibitemShut {NoStop}%
\bibitem [{\citenamefont {Reimer}\ \emph {et~al.}(2019)\citenamefont {Reimer},
  \citenamefont {Hu}, \citenamefont {Shams-Ansari}, \citenamefont {Zhang},\
  and\ \citenamefont {Loncar}}]{Reimer:1909.01303:ARXIV}%
  \BibitemOpen
  \bibfield  {author} {\bibinfo {author} {\bibfnamefont {C.}~\bibnamefont
  {Reimer}}, \bibinfo {author} {\bibfnamefont {Y.}~\bibnamefont {Hu}}, \bibinfo
  {author} {\bibfnamefont {A.}~\bibnamefont {Shams-Ansari}}, \bibinfo {author}
  {\bibfnamefont {M.}~\bibnamefont {Zhang}}, \ and\ \bibinfo {author}
  {\bibfnamefont {M.}~\bibnamefont {Loncar}},\ }\bibfield  {title} {\enquote
  {\bibinfo {title} {{High-dimensional frequency crystals and quantum walks in
  electro-optic microcombs}},}\ }\href {http://arxiv.org/abs/1909.01303}
  {\bibfield  {journal} {\bibinfo  {journal} {arXiv}\ }\textbf {\bibinfo
  {volume} {\mdseries 1909.01303}} (\bibinfo {year} {2019})}\BibitemShut
  {NoStop}%
\bibitem [{\citenamefont {Bersch}\ \emph {et~al.}(2009)\citenamefont {Bersch},
  \citenamefont {Onishchukov},\ and\ \citenamefont
  {Peschel}}]{Bersch:2009-2372:OL}%
  \BibitemOpen
  \bibfield  {author} {\bibinfo {author} {\bibfnamefont {C.}~\bibnamefont
  {Bersch}}, \bibinfo {author} {\bibfnamefont {G.}~\bibnamefont {Onishchukov}},
  \ and\ \bibinfo {author} {\bibfnamefont {U.}~\bibnamefont {Peschel}},\
  }\bibfield  {title} {{\selectlanguage {English}\enquote {\bibinfo {title}
  {Experimental observation of spectral {B}loch oscillations},}\ }}\href
  {https://doi.org/10.1364/OL.34.002372} {\bibfield  {journal} {\bibinfo
  {journal} {Opt. Lett.}\ }\textbf {\bibinfo {volume} {34}},\ \bibinfo {pages}
  {2372--2374} (\bibinfo {year} {2009})}\BibitemShut {NoStop}%
\bibitem [{\citenamefont {Kang}\ \emph {et~al.}(2009)\citenamefont {Kang},
  \citenamefont {Nazarkin}, \citenamefont {Brenn},\ and\ \citenamefont
  {Russell}}]{Kang:2009-276:NPHYS}%
  \BibitemOpen
  \bibfield  {author} {\bibinfo {author} {\bibfnamefont {M.~S.}\ \bibnamefont
  {Kang}}, \bibinfo {author} {\bibfnamefont {A.}~\bibnamefont {Nazarkin}},
  \bibinfo {author} {\bibfnamefont {A.}~\bibnamefont {Brenn}}, \ and\ \bibinfo
  {author} {\bibfnamefont {P.~S.~J.}\ \bibnamefont {Russell}},\ }\bibfield
  {title} {{\selectlanguage {English}\enquote {\bibinfo {title} {Tightly
  trapped acoustic phonons in photonic crystal fibres as highly nonlinear
  artificial {R}aman oscillators},}\ }}\href
  {https://doi.org/10.1038/NPHYS1217} {\bibfield  {journal} {\bibinfo
  {journal} {Nat. Phys.}\ }\textbf {\bibinfo {volume} {5}},\ \bibinfo {pages}
  {276--280} (\bibinfo {year} {2009})}\BibitemShut {NoStop}%
\bibitem [{\citenamefont {Wolff}\ \emph {et~al.}(2017)\citenamefont {Wolff},
  \citenamefont {Stiller}, \citenamefont {Eggleton}, \citenamefont {Steel},\
  and\ \citenamefont {Poulton}}]{Wolff:2017-23021:NJP}%
  \BibitemOpen
  \bibfield  {author} {\bibinfo {author} {\bibfnamefont {C.}~\bibnamefont
  {Wolff}}, \bibinfo {author} {\bibfnamefont {B.}~\bibnamefont {Stiller}},
  \bibinfo {author} {\bibfnamefont {B.~J.}\ \bibnamefont {Eggleton}}, \bibinfo
  {author} {\bibfnamefont {M.~J.}\ \bibnamefont {Steel}}, \ and\ \bibinfo
  {author} {\bibfnamefont {C.~G.}\ \bibnamefont {Poulton}},\ }\bibfield
  {title} {{\selectlanguage {English}\enquote {\bibinfo {title} {Cascaded
  forward {B}rillouin scattering to all {S}tokes orders},}\ }}\href
  {https://doi.org/10.1088/1367-2630/aa599e} {\bibfield  {journal} {\bibinfo
  {journal} {New J. Phys.}\ }\textbf {\bibinfo {volume} {19}},\ \bibinfo
  {pages} {023021} (\bibinfo {year} {2017})}\BibitemShut {NoStop}%
\bibitem [{\citenamefont {Eggleton}\ \emph {et~al.}(2019)\citenamefont
  {Eggleton}, \citenamefont {Poulton}, \citenamefont {Rakich}, \citenamefont
  {Steel},\ and\ \citenamefont {Bahl}}]{Eggleton:2019-664:NPHOT}%
  \BibitemOpen
  \bibfield  {author} {\bibinfo {author} {\bibfnamefont {B.~J.}\ \bibnamefont
  {Eggleton}}, \bibinfo {author} {\bibfnamefont {C.~G.}\ \bibnamefont
  {Poulton}}, \bibinfo {author} {\bibfnamefont {P.~T.}\ \bibnamefont {Rakich}},
  \bibinfo {author} {\bibfnamefont {M.~J.}\ \bibnamefont {Steel}}, \ and\
  \bibinfo {author} {\bibfnamefont {G.}~\bibnamefont {Bahl}},\ }\bibfield
  {title} {{\selectlanguage {English}\enquote {\bibinfo {title} {Brillouin
  integrated photonics},}\ }}\href {https://doi.org/10.1038/s41566-019-0498-z}
  {\bibfield  {journal} {\bibinfo  {journal} {Nat. Photon.}\ }\textbf {\bibinfo
  {volume} {13}},\ \bibinfo {pages} {664--677} (\bibinfo {year}
  {2019})}\BibitemShut {NoStop}%
\bibitem [{\citenamefont {Reimer}\ \emph {et~al.}(2016)\citenamefont {Reimer},
  \citenamefont {Kues}, \citenamefont {Roztocki}, \citenamefont {Wetzel},
  \citenamefont {Grazioso}, \citenamefont {Little}, \citenamefont {Chu},
  \citenamefont {Johnston}, \citenamefont {Bromberg}, \citenamefont {Caspani},
  \citenamefont {Moss},\ and\ \citenamefont
  {Morandotti}}]{Reimer:2016-1176:SCI}%
  \BibitemOpen
  \bibfield  {author} {\bibinfo {author} {\bibfnamefont {C.}~\bibnamefont
  {Reimer}}, \bibinfo {author} {\bibfnamefont {M.}~\bibnamefont {Kues}},
  \bibinfo {author} {\bibfnamefont {P.}~\bibnamefont {Roztocki}}, \bibinfo
  {author} {\bibfnamefont {B.}~\bibnamefont {Wetzel}}, \bibinfo {author}
  {\bibfnamefont {F.}~\bibnamefont {Grazioso}}, \bibinfo {author}
  {\bibfnamefont {B.~E.}\ \bibnamefont {Little}}, \bibinfo {author}
  {\bibfnamefont {S.~T.}\ \bibnamefont {Chu}}, \bibinfo {author} {\bibfnamefont
  {T.}~\bibnamefont {Johnston}}, \bibinfo {author} {\bibfnamefont
  {Y.}~\bibnamefont {Bromberg}}, \bibinfo {author} {\bibfnamefont
  {L.}~\bibnamefont {Caspani}}, \bibinfo {author} {\bibfnamefont {D.~J.}\
  \bibnamefont {Moss}}, \ and\ \bibinfo {author} {\bibfnamefont
  {R.}~\bibnamefont {Morandotti}},\ }\bibfield  {title} {{\selectlanguage
  {English}\enquote {\bibinfo {title} {Generation of multiphoton entangled
  quantum states by means of integrated frequency combs},}\ }}\href
  {https://doi.org/10.1126/science.aad8532} {\bibfield  {journal} {\bibinfo
  {journal} {Science}\ }\textbf {\bibinfo {volume} {351}},\ \bibinfo {pages}
  {1176--1180} (\bibinfo {year} {2016})}\BibitemShut {NoStop}%
\bibitem [{\citenamefont {Perets}\ \emph {et~al.}(2008)\citenamefont {Perets},
  \citenamefont {Lahini}, \citenamefont {Pozzi}, \citenamefont {Sorel},
  \citenamefont {Morandotti},\ and\ \citenamefont
  {Silberberg}}]{Perets:2008-170506:PRL}%
  \BibitemOpen
  \bibfield  {author} {\bibinfo {author} {\bibfnamefont {H.~B.}\ \bibnamefont
  {Perets}}, \bibinfo {author} {\bibfnamefont {Y.}~\bibnamefont {Lahini}},
  \bibinfo {author} {\bibfnamefont {F.}~\bibnamefont {Pozzi}}, \bibinfo
  {author} {\bibfnamefont {M.}~\bibnamefont {Sorel}}, \bibinfo {author}
  {\bibfnamefont {R.}~\bibnamefont {Morandotti}}, \ and\ \bibinfo {author}
  {\bibfnamefont {Y.}~\bibnamefont {Silberberg}},\ }\bibfield  {title}
  {{\selectlanguage {English}\enquote {\bibinfo {title} {Realization of quantum
  walks with negligible decoherence in waveguide lattices},}\ }}\href
  {https://doi.org/10.1103/PhysRevLett.100.170506} {\bibfield  {journal}
  {\bibinfo  {journal} {Phys. Rev. Lett.}\ }\textbf {\bibinfo {volume} {100}},\
  \bibinfo {pages} {170506} (\bibinfo {year} {2008})}\BibitemShut {NoStop}%
\bibitem [{\citenamefont {Thurston}(1997)}]{Thurston:1997:ThreeDimensional}%
  \BibitemOpen
  \bibfield  {author} {\bibinfo {author} {\bibfnamefont {W.~P.}\ \bibnamefont
  {Thurston}},\ }\href
  {https://press.princeton.edu/books/hardcover/9780691083049/three-dimensional-geometry-and-topology-volume-1}
  {\emph {\bibinfo {title} {Three-dimensional geometry and topology}}},\
  Vol.~\bibinfo {volume} {35}\ (\bibinfo  {publisher} {Princeton University
  Press},\ \bibinfo {year} {1997})\BibitemShut {NoStop}%
\bibitem [{\citenamefont {Weeks}(2001)}]{Weeks:2001:ShapeSpace}%
  \BibitemOpen
  \bibfield  {author} {\bibinfo {author} {\bibfnamefont {J.~R.}\ \bibnamefont
  {Weeks}},\ }\href
  {https://www.crcpress.com/The-Shape-of-Space/Weeks/p/book/9780824707095}
  {\emph {\bibinfo {title} {The shape of space}}}\ (\bibinfo  {publisher} {CRC
  press},\ \bibinfo {year} {2001})\BibitemShut {NoStop}%
\bibitem [{\citenamefont {Eichelkraut}\ \emph {et~al.}(2014)\citenamefont
  {Eichelkraut}, \citenamefont {Vetter}, \citenamefont {Perez-Leija},
  \citenamefont {Moya-Cessa}, \citenamefont {Christodoulides},\ and\
  \citenamefont {Szameit}}]{Eichelkraut:2014-268:OPT}%
  \BibitemOpen
  \bibfield  {author} {\bibinfo {author} {\bibfnamefont {T.}~\bibnamefont
  {Eichelkraut}}, \bibinfo {author} {\bibfnamefont {C.}~\bibnamefont {Vetter}},
  \bibinfo {author} {\bibfnamefont {A.}~\bibnamefont {Perez-Leija}}, \bibinfo
  {author} {\bibfnamefont {H.}~\bibnamefont {Moya-Cessa}}, \bibinfo {author}
  {\bibfnamefont {D.~N.}\ \bibnamefont {Christodoulides}}, \ and\ \bibinfo
  {author} {\bibfnamefont {A.}~\bibnamefont {Szameit}},\ }\bibfield  {title}
  {{\selectlanguage {English}\enquote {\bibinfo {title} {Coherent random walks
  in free space},}\ }}\href {https://doi.org/10.1364/OPTICA.1.000268}
  {\bibfield  {journal} {\bibinfo  {journal} {Optica}\ }\textbf {\bibinfo
  {volume} {1}},\ \bibinfo {pages} {268--271} (\bibinfo {year}
  {2014})}\BibitemShut {NoStop}%
\bibitem [{\citenamefont {Titchener}\ \emph {et~al.}(2020)\citenamefont
  {Titchener}, \citenamefont {Bell}, \citenamefont {Wang}, \citenamefont
  {Solntsev}, \citenamefont {Eggleton},\ and\ \citenamefont
  {Sukhorukov}}]{Wang:2002.09160:ARXIV}%
  \BibitemOpen
  \bibfield  {author} {\bibinfo {author} {\bibfnamefont {J.~G.}\ \bibnamefont
  {Titchener}}, \bibinfo {author} {\bibfnamefont {B.}~\bibnamefont {Bell}},
  \bibinfo {author} {\bibfnamefont {K.}~\bibnamefont {Wang}}, \bibinfo {author}
  {\bibfnamefont {A.~S.}\ \bibnamefont {Solntsev}}, \bibinfo {author}
  {\bibfnamefont {B.~J.}\ \bibnamefont {Eggleton}}, \ and\ \bibinfo {author}
  {\bibfnamefont {A.~A.}\ \bibnamefont {Sukhorukov}},\ }\bibfield  {title}
  {\enquote {\bibinfo {title} {{Synthetic photonic lattice for single-shot
  reconstruction of frequency combs}},}\ }\href
  {https://arxiv.org/abs/2002.09160} {\bibfield  {journal} {\bibinfo  {journal}
  {arXiv}\ }\textbf {\bibinfo {volume} {\mdseries 2002.09160}} (\bibinfo {year}
  {2020})}\BibitemShut {NoStop}%
\bibitem [{\citenamefont {Wang}\ \emph {et~al.}(2017)\citenamefont {Wang},
  \citenamefont {Shi}, \citenamefont {Solntsev}, \citenamefont {Fan},
  \citenamefont {Sukhorukov},\ and\ \citenamefont
  {Neshev}}]{Wang:2017-1990:OL}%
  \BibitemOpen
  \bibfield  {author} {\bibinfo {author} {\bibfnamefont {K.}~\bibnamefont
  {Wang}}, \bibinfo {author} {\bibfnamefont {Y.}~\bibnamefont {Shi}}, \bibinfo
  {author} {\bibfnamefont {A.~S.}\ \bibnamefont {Solntsev}}, \bibinfo {author}
  {\bibfnamefont {S.~H.}\ \bibnamefont {Fan}}, \bibinfo {author} {\bibfnamefont
  {A.~A.}\ \bibnamefont {Sukhorukov}}, \ and\ \bibinfo {author} {\bibfnamefont
  {D.~N.}\ \bibnamefont {Neshev}},\ }\bibfield  {title} {{\selectlanguage
  {English}\enquote {\bibinfo {title} {Non-reciprocal geometric phase in
  nonlinear frequency conversion},}\ }}\href
  {https://doi.org/10.1364/OL.42.001990} {\bibfield  {journal} {\bibinfo
  {journal} {Opt. Lett.}\ }\textbf {\bibinfo {volume} {42}},\ \bibinfo {pages}
  {1990--1993} (\bibinfo {year} {2017})}\BibitemShut {NoStop}%
\bibitem [{\citenamefont {Shi}\ \emph {et~al.}(2015)\citenamefont {Shi},
  \citenamefont {Yu},\ and\ \citenamefont {Fan}}]{Shi:2015-388:NPHOT}%
  \BibitemOpen
  \bibfield  {author} {\bibinfo {author} {\bibfnamefont {Y.}~\bibnamefont
  {Shi}}, \bibinfo {author} {\bibfnamefont {Z.~F.}\ \bibnamefont {Yu}}, \ and\
  \bibinfo {author} {\bibfnamefont {S.~H.}\ \bibnamefont {Fan}},\ }\bibfield
  {title} {{\selectlanguage {English}\enquote {\bibinfo {title} {Limitations of
  nonlinear optical isolators due to dynamic reciprocity},}\ }}\href
  {https://doi.org/10.1038/NPHOTON.2015.79} {\bibfield  {journal} {\bibinfo
  {journal} {Nat. Photon.}\ }\textbf {\bibinfo {volume} {9}},\ \bibinfo {pages}
  {388--392} (\bibinfo {year} {2015})}\BibitemShut {NoStop}%
\bibitem [{\citenamefont {Jukic}\ and\ \citenamefont
  {Buljan}(2013)}]{Jukic:2013-13814:PRA}%
  \BibitemOpen
  \bibfield  {author} {\bibinfo {author} {\bibfnamefont {D.}~\bibnamefont
  {Jukic}}\ and\ \bibinfo {author} {\bibfnamefont {H.}~\bibnamefont {Buljan}},\
  }\bibfield  {title} {{\selectlanguage {English}\enquote {\bibinfo {title}
  {Four-dimensional photonic lattices and discrete tesseract solitons},}\
  }}\href {https://doi.org/10.1103/PhysRevA.87.013814} {\bibfield  {journal}
  {\bibinfo  {journal} {Phys. Rev. A}\ }\textbf {\bibinfo {volume} {87}},\
  \bibinfo {pages} {013814} (\bibinfo {year} {2013})}\BibitemShut {NoStop}%
\bibitem [{\citenamefont {Fang}\ \emph {et~al.}(2012)\citenamefont {Fang},
  \citenamefont {Yu},\ and\ \citenamefont {Fan}}]{Fang:2012-782:NPHOT}%
  \BibitemOpen
  \bibfield  {author} {\bibinfo {author} {\bibfnamefont {K.~J.}\ \bibnamefont
  {Fang}}, \bibinfo {author} {\bibfnamefont {Z.~F.}\ \bibnamefont {Yu}}, \ and\
  \bibinfo {author} {\bibfnamefont {S.~H.}\ \bibnamefont {Fan}},\ }\bibfield
  {title} {{\selectlanguage {English}\enquote {\bibinfo {title} {Realizing
  effective magnetic field for photons by controlling the phase of dynamic
  modulation},}\ }}\href {https://doi.org/10.1038/NPHOTON.2012.236} {\bibfield
  {journal} {\bibinfo  {journal} {Nat. Photon.}\ }\textbf {\bibinfo {volume}
  {6}},\ \bibinfo {pages} {782--787} (\bibinfo {year} {2012})}\BibitemShut
  {NoStop}%
\bibitem [{\citenamefont {Longhi}(2015)}]{Longhi:2015-2941:OL}%
  \BibitemOpen
  \bibfield  {author} {\bibinfo {author} {\bibfnamefont {S.}~\bibnamefont
  {Longhi}},\ }\bibfield  {title} {{\selectlanguage {English}\enquote {\bibinfo
  {title} {Synthetic gauge fields for light beams in optical resonators},}\
  }}\href {https://doi.org/10.1364/OL.40.002941} {\bibfield  {journal}
  {\bibinfo  {journal} {Opt. Lett.}\ }\textbf {\bibinfo {volume} {40}},\
  \bibinfo {pages} {2941--2944} (\bibinfo {year} {2015})}\BibitemShut {NoStop}%
\end{thebibliography}
\end{document}